\documentclass[10pt]{article}
\usepackage[top=0.85in,left=2.75in,footskip=0.75in,marginparwidth=2in]{geometry}

\usepackage[utf8]{inputenc}

\usepackage{cite}
\usepackage{amsmath}
\usepackage{nameref,hyperref}

\usepackage[right]{lineno}

\usepackage{microtype}
\DisableLigatures[f]{encoding = *, family = * }

\raggedright
\usepackage{changepage}

\usepackage[aboveskip=1pt,labelfont=bf,labelsep=period,singlelinecheck=off]{caption}

\makeatletter
\renewcommand{\@biblabel}[1]{\quad#1.}
\makeatother

\usepackage{lastpage,fancyhdr,graphicx}
\usepackage{epstopdf}
\fancyheadoffset[L]{2.25in}
\fancyfootoffset[L]{2.25in}

\usepackage{color}

\definecolor{Gray}{gray}{.25}

\usepackage{graphicx}

\usepackage{sidecap}

\usepackage{wrapfig}
\usepackage[pscoord]{eso-pic}
\usepackage[fulladjust]{marginnote}
\reversemarginpar

\begin{document}
\vspace*{0.35in}

\begin{flushleft}
{\Large
\textbf\newline{Statistical characterization and time-series modeling of seismic noise}
}
\newline
\\
Kanchan Aggarwal\textsuperscript{1},
Siddhartha Mukhopadhyay\textsuperscript{2},
Arun K Tangirala \textsuperscript{1,*}
\\
\bigskip
\bf{1} Department of Chemical Engineering, Indian Institute of Technology Madras, Tamil Nadu, India
\\
\bf{2} Seismology Division,	Bhabha Atomic Research Centre, Mumbai, India
\\
\bigskip
* arunkt@iitm.ac.in

\end{flushleft}

\section*{Abstract}
Developing statistical models for seismic noise is an exercise of high value in seismic data analysis since these models play a critical role in detecting the onset of seismic events. A majority of these models are usually built on certain critical assumptions, namely, stationarity, linearity, and Gaussianity. Despite their criticality, very little reported literature exists on validating these assumptions on real seismic data. The objectives of this work are (i) to critically study these long-held assumptions and (ii) to propose a systematic procedure for developing appropriate time-series models. A rigorous statistical analysis reveals that these standard assumptions do not hold for most of the data sets under study; rather they exhibit additional special features such as heteroskedasticity and integrating effects. Resting on these novel discoveries, ARIMA-GARCH models are developed for seismic noise. Studies are carried out on $185$ real-time data sets over different time intervals to study the daily and seasonal variations in noise characteristics and model structure. Nearly all datasets tested positive for first-order non-stationarity, heteroskedasticity, and Gaussianity, while $19\%$ tested negative for linearity. Analysis of the structural uniformity of the developed models with respect to daily and seasonal variations is also presented.


\section*{Introduction}
Analysis of seismic or under-the-ground data has enormous significance for obvious important reasons. The ground is at unrest continuously mainly due to oceanic waves, changes in earth’s crust, atmospheric variations, and human activities. The primary goal of seismic data analysis is to develop tools for reliable earthquakes detection systems. These tools involve onset detection of events, source location estimation, identification of underground nuclear explosions, imaging of Earth's deep interior structures (tomography), prediction of earthquakes using data mining techniques and preparing the seismic risk maps for highly earthquake-prone regions \cite{zakeri2015data}. The major challenge in seismic data analysis is to uncover the primary events in the presence of noise, especially from low signal to noise ratio (SNR) measurements. In seismic literature, seismic noise is perceived as any undesirable signal recorded by a seismometer in the absence of a seismic event. The success of any seismic data analytic method, therefore, rests on an appropriate characterization and modeling of seismic noise, which is the focus of this study.

A majority of the seismic applications assume that seismic noise possesses certain properties such as stationary or locally stationary and that it is linear and driven by Gaussian distributed shock waves (\cite{wang1999random}, \cite{wang2009multisensor}, \cite{jiang2010directional}, \cite{wang2002seismic}). Contrary to these assumptions, seismic noise may possess different characteristics that vary with space and time depending on various natural and human-induced factors. Therefore, one should expect that these assumptions that are usually made for mathematical convenience, may not tally with the physical nature of seismic data. Despite these intuitive expectations, there have been very scarce efforts in the reported literature towards a rigorous study for verifying the veracity of these assumptions. Not only this study is of importance from a rigour standpoint but also from the viewpoint of selecting the appropriate model class during the model development stage. Thus, a logical procedure for model development should first involve proper characterization of noise followed by a choice of a model class that is commensurate with the discovered noise characteristics.

The overall objectives of this work, motivated by the foregoing arguments, are (i) to conduct a rigorous study for systematically characterizing the seismic noise and (ii) to build a suitable time-series model based on the discovered noise characteristics from the preceding step. The focus of this work is on verifying four prominent characteristics of noise, namely, integrating (random walk) effects, \emph{heteroskedasticity}, linearity, and Gaussianity. The first two characteristics pertain to two specific types of non-stationarities, while the remaining two are related to the model structure and distribution of driving white noise. All four factors affect the optimality of prediction in their own right. A warranted discussion of this effect is provided in the following paragraphs.

In principle, there exist different types of non-stationarities such as integrating effects, deterministic trends (a polynomial function of time including seasonality or periodicity),  heteroskedasticity (changing variance), time-varying higher-order moments, etc. However, we restrict our quest to only the two aforementioned types for two reasons. The first reason is from a process viewpoint, that is, we expect seismic data to contain these characteristics due to the nature of seismic wave generation processes and /or the instrumentation devices associated with measurements. Turning to the second characteristic of interest, it is natural that factors contributing to seismic noise can change with time, thereby leading to time-varying statistical properties, second-order moments to be specific. This changing variability, known as \emph{heteroskedasticity}, can also occur due to the change in properties of the seismic wave as it travels through the different layers of Earth.

The second reason for studying the two statistical properties of interest is from a remedy or solution viewpoint. An extensive body of literature exists on modeling these effects owing to the fact that random walk and heteroskedasticity are commonly encountered in stochastic processes of other fields such as econometrics, meteorology, climatology, and engineering (\cite{modarres2014modeling},\cite{CAPORIN20123459}). Integrating effects are modeled as ARIMA (auto-regressive Integrated Moving Average) models. On the other hand, models that capture heteroskedasticity depend on the type, i.e., whether the series exhibits \emph{unconditional} or \emph{conditional} heteroskedasticity. The former class of processes are modeled by suitable transformations such as Box-Cox type (\cite{sakia1992box}), while the latter class requires a more complex model structure.
Engle provided a modeling framework for such series \cite{engle1982autoregressive}, which is known under the name of ARCH (auto-regressive conditionally heteroskedastic) models. \ref{sec:tsmodel} and \ref{subsec:ARCHmodel} briefly review the mathematical descriptions associated with ARIMA and GARCH (Generalized ARCH) models.

Properties such as linearity and Gaussianity are important from the optimality of predictions viewpoint. Subject to the condition of stationarity, a \emph{linear} time-series model (essentially an ARIMA model) results in optimal predictions (among all classes of models) when the data generating process is jointly Gaussian. A random process can be linear and non-Gaussian or be the converse, non-linear and Gaussian (see \ref{sec:LinearityTest} and \ref{sec:GaussianityTest}  for technical definitions of linearity and Gaussianity). Linearity and Gaussianity, although closely related are different properties. When one of these two assumptions is violated, the ARIMA class of models result in sub-optimal predictions, which in turn implies incomplete capture of process characteristics. The impact is then manifest in the success rate of event detection technique, especially if it is based on model predictions. Therefore, it is always advisable to test the series for these properties rather than assuming them to be holding true. 

The foregoing discussion provides a strong impetus for a careful study of seismic noise and hence its modeling. The related literature however is, interestingly, devoid of such studies barring a handful of works. Recently, \cite{Zhong2014}, \cite{zhong2015statistical} and \cite{wang2014study} have drawn attention towards noise characterization based on the aforementioned properties. Zhong et al. ($2014$ and $2015$) \cite{Zhong2014,zhong2015statistical} characterized the noise using a surrogate-based time-frequency approach (\cite{DSouza}, \cite{Brognat}) for stationarity. However, there is no specific mention of the type of non-stationarity that the series is being tested for. There is no attempt or discussion in the recorded literature on possible heteroskedastic effects in seismic noise. With regards to linearity and Gaussianity, a bi-spectral based method \cite{hinich1982testing}, \cite{BIRKELUND20092537} is implemented. Wang et al. \cite{wang2014study} uses Shapiro-Wilk test \cite{shapiro1965analysis} along with the surrogate and time-frequency based approach for testing stationarity. A technical lacuna in using the Shapiro-Wilk (and any other Gaussianity) test is that they are suited only for weakly correlated data, which is not the case with seismic noise. Therefore, we choose to implement the test on residuals or prediction errors generated by the linear time-series model while the a modern surrogate-based test is deployed for testing linearity. The premise is that since the white residual series is the forcing function for the linear model, its Gaussianity implies that the given seismic series is also jointly Gaussian. With regards to stationarity tests in this work, we deploy simple statistical tests such as Augmented Dickey Fuller (ADF) and Phillips Perron (PP) test for testing the presence of integrating effect, while Priestley Subba Rao (PSR) and autoregressive conditional heteroskedastic (ARCH)  test for testing the heteroskedasticity.

The primary findings of our study are two-fold. Firstly, that seismic noise exhibits (conditional) heteroskedasticity and integrating effects, contrary to the standard stationarity assumptions or in the least those not speculated to be present in seismic data. These constitute the first main contribution of the present work. Secondly, series across most stations are tested to be positive for linear and driven by Gaussian uncorrelated signals, thereby providing sufficient support for the widespread use of linear models in seismic data analysis. However, the first set of findings call for the development of ARIMA-GARCH models, which constitute the latter half of this work. Furthermore, we also study the daily and seasonal variations in noise properties and model structure. The case studies reveal that the noise properties are invariant on daily and seasonal basis. However, estimated model parameters are found to vary with time, which is meaningful given that sources of seismic noise (e.g., human-induced contributions) are likely to vary in a given 24-hour period.

The rest of the article is as follows. In Section \ref{sec:Proposedmethod}, we describe the proposed systematic method for characterizing the seismic noise through statistical hypothesis tests and developing a suitable time-series model based on the outcomes of hypothesis tests. Analysis of $185$ datasets collected from ANMO station using the proposed method is presented in Section \ref{sec:Results}. The paper ends with a few concluding remarks in Section \ref{sec:Conclusions}.

\section{Proposed systematic methodology}\label{sec:Proposedmethod}

\begin{figure}[t]
	\centering
	\includegraphics[scale=0.35]{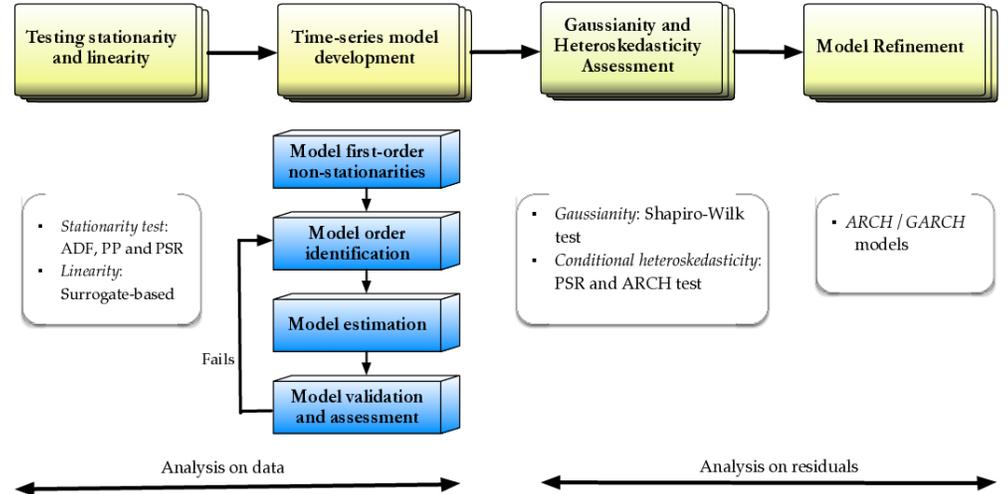}
	\caption{Proposed methodology\label{Fig:methodology}}
\end{figure}

This section presents the main contribution of this work, where a systematic methodology (as shown in Figure \ref{Fig:methodology}) for the statistical characterization of seismic noise is presented. The procedure is divided into two parts, (i) analysis of the \emph{given data} and (ii) analysis of the \emph{pre-whitened data}. Details of different steps involved in each part are discussed in the following sections.

\subsection{Testing Stationarity and Linearity}\label{subsec:Step1}
Stationarity and linearity analysis are conducted directly on data. The sequence in which these tests are performed is essential because the linearity test assumes that the data is first-order stationary.

\textbf{Stationarity analysis:} The first step in characterizing the seismic noise is to test the data for the presence of first-order non-stationarities (integrating effects or trends). It involves a combination of visual and statistical analysis to make qualitative and quantitative inferences, respectively. Integrating processes are first-order auto-regressive (AR) process with a characteristic root at unity. Therefore, integrating effects are also called as \emph{unit root effects}.

(i) \textit{Visual analysis:} The presence of such non-stationarities is assessed by visual inspection of the autocorrelation function (ACF) and partial-correlation function (PACF) of the noise. A slowly decaying nature of ACF indicates the presence of deterministic trends or integrating effects in data. Both the non-stationarities result in significant low-frequency or highly correlated data. Their presence can also be inferred from the near-unity value of PACF at initial lags other than zero; the number of such lags depends on the order of integrating effects. 

(ii) \textit{Statistical analysis:} Inferences drawn from the visual analysis are quantitatively verified by statistical unit root tests such as ADF and PP tests. In the presence of heteroskedasticity, the test statistics for ADF and PP decreases in value, resulting in increased false rejection of the null hypothesis that unit root is present in data \cite{cavaliere2005unit}. For heteroskedastic data, these tests are not guaranteed to be effective \cite{WESTERLUND201440}. Therefore, we propose an alternative but less rigorous method. We fit an AR($1$) model to the data and analyze the estimated coefficient. A near-unity value of the AR($1$) coefficient, a value higher than $0.967$, confirms the presence of integrating effects. This approach is suited for all situations regardless of whether the series is heteroskedastic or otherwise.

If the series is being tested positive for first-order non-stationarities, then these effects are modeled prior to the linearity test. Trend type non-stationarity is modeled either by the polynomial fitting of a suitable order or through suitable filtering/smoothing techniques \cite{brockwell2002introduction}. Integrating type non-stationarity is modeled by differencing the data up to a suitable degree.

\textbf{Linearity analysis:} After the non-stationarity effects are modeled, the ``stationarized`` data is then tested for linearity using a surrogate-based hypothesis test. In recent years, the surrogate-based approach has emerged as a powerful method over the traditional non-linearity tests such as time-reversibility test, BDS test, etc., \cite{tsay1986nonlinearity,theiler1992testing} for two key reasons. Firstly, the test can be tailored to a specific null hypothesis. Secondly,  it is compatible with any test statistic that can be selected independently of the null hypothesis. In this work, surrogates are generated using the amplitude-adjusted Fourier transform (AAFT) algorithm (\cite{schreiber2000surrogate}) and \textit{correlation dimension}  \cite{grassberger1983measuring} is used as the test statistic for conducting the test. The correlation dimension ($D_{2}$) is a measure of dimensionality of the space occupied by the random data points. At a chosen significance level, if $D_{2}$ of data differs from the ensemble of surrogates, then the null hypothesis that the data is generated from a linear Gaussian process is rejected in favor of the alternate hypothesis.

\subsection{Model development}\label{subsec:Step2}
Post characterization and any pre-treatment of data, the next step is to develop a time-series model that is \emph{commensurate} with the noise properties. In the first stage of modeling, we develop an optimal ARIMA($p,d,m$) models of suitable order following the procedure outlined below. The procedure given below is usually standard and widely available in the literature (\cite{box2015time,brockwell2002introduction}). We outline them for completeness sake.\\
(i) \textit{Modeling integrating effects:} The series, if tested positive for presence of integrating effects or random walk behaviour, is differenced a suitable number of times, until the series tests negative. This degree of differencing is determined by a repeated differencing of the series and running the statistical tests at each stage of differencing, until the null hypothesis is rejected.\\
(ii) \textit{Estimate the model of suitable order:} Stationary time-series is then modeled using ARMA($p,m$) model, where the $p+m$ parameters and the noise variance are estimated using estimation methods such as least squares, maximum likelihood, etc. Suitable initial guesses of model orders $p$ and $m$ are made through an examination of the autocorrelation function (ACF) and partial autocorrelation function (PACF) of data.\\
(iii) \textit{Model assessment and validation:} The goodness of the estimated model is assessed through a statistical analysis of residuals for \emph{underfitting}. Specifically, the residuals are tested for whiteness through a significance test on the ACFs or a Box-Ljung-Pierce test. In this work, we use the former approach. If the residuals test negative for whiteness, the model order is refined until a satisfactory result is obtained. Once the model passes the test of underfitting, the parameter estimates are subjected to a significance test so as to check for \emph{overfitting}. At a significance level $\alpha$, the $100(1-\alpha)\%$ confidence interval for all parameters should exclude zero for the null hypothesis (that the true parameter is zero) is rejected. If any of the parameter estimate(s) is found to be insignificant, the model is re-estimated after removing the corresponding term(s) in the difference equation. The resulting model is again tested for underfit and overfit. This procedure is repeated until a satisfactory  model is obtained.

It may be noted that information-theoretic criteria such as Akaike information criterion (AIC), Bayesian / Schwartz information criterion (BIC) and minimum description length (MDL) are widely used for the determination of suitable model orders. However, it is important that these approaches usually result in good final guesses, but not necessarily the most appropriate order. Models obtained through such approaches also have to be subject to the tests of underfit and overfit as described above.

\subsection{Assessment of Gaussianity and Heteroskedasticity}\label{subsec:Step3}
Statistical tests for Gaussianity and conditional heteroskedasticity, unlike the stationarity and linearity tests, require the data to be uncorrelated or weakly correlated. Therefore, these tests are best implemented on pre-whitened data. Furthermore, the statistical test for heteroskedasticity assumes that data is generated from a Gaussian random process. It is, therefore, necessary to test the pre-whitened data for normality before testing for heteroskedasticity.

\textbf{Gaussianity analysis:} In this work, Shapiro-Wilk test \cite{shapiro1965analysis} is used to test the normality of pre-whitened. The null hypothesis that a linear Gaussian process drives the data is rejected in favor of the alternate if the $p$-value is smaller than the chosen significance level. The test is implemented on $2000$ data points because the performance of SW test is limited by the sample size ($<5000$).

\textbf{Heteroskedasticity analysis:} The Presence of heteroskedasticity is tested through visual analysis of squared pre-whitened data followed by two statistical tests, namely, PSR and ARCH. A significant correlation in the squared pre-whitened data indicates the presence of conditional heteroskedasticity \cite{mcleod1983diagnostic}. This effect is also known as the ARCH effect. Qualitative inferences drawn from the visual analysis are quantitatively verified using PSR and ARCH tests.

\section{Results and Discussions}\label{sec:Results} We present two case studies to demonstrate the utility of the proposed methodology to characterize and model the seismic noise. In the first case study, daily variations in the noise properties and hence the model structure are studied for a fixed geographical location. In the second case study, the uniqueness and uniformity of the model structure along with the statistical properties are studied for monthly variations.
\begin{table}[t]
	\centering
	\caption{Specifications of datasets for all the case studies downloaded from IRIS.\label{tab1:spec}}
	\centering\resizebox*{\linewidth}{!}{
		\begin{tabular}{cccccccc}
			\hline
			\textbf{Data} & \multicolumn{1}{l}{\textbf{Network}} & \multicolumn{1}{l}{\textbf{Station}} & \textbf{Channel} & \textbf{Location} & \textbf{Date} & \textbf{\begin{tabular}[c]{@{}l@{}}Start time\\ (UTC)\end{tabular}} & \textbf{\begin{tabular}[c]{@{}l@{}}End time\\ (UTC)\end{tabular}} \\ \hline
			1. & IU & ANMO & BHZ (20 sps) & US & 27.02.2010 & 17:30 & 18:30 \\
			2. & IU & ANMO & BHZ (20 sps) & US & 01.03.2010 & 22:00 & 23:00 \\
			3. & IU & ANMO & BHZ (20 sps) & US & 02.03.2010 & 22:00 & 23:00 \\
			: & : & : & : & : & : & : & : \\
			61. & IU & ANMO & BHZ (20 sps) & US & 29.03.2010 & 22:00 & 23:00 \\
			62. & IU & ANMO & BHZ (20 sps) & US & 30.04.2010 & 22:00 & 23:00 \\
			63. & IU & ANMO & BHZ (20 sps) & US & 01.12.2010 & 22:00 & 23:00 \\
			64. & IU & ANMO & BHZ (20 sps) & US & 02.12.2010 & 22:00 & 23:00 \\
			: & : &  & : & : & : & : & : \\
			92. & IU & ANMO & BHZ (20 sps) & US & 30.12.2010 & 22:00 & 23:00 \\
			93. & IU & ANMO & BHZ (20 sps) & US & 31.12.2010 & 22:00 & 23:00 \\
			94. & IU & ANMO & BHZ (20 sps) & US & 01.03.2010 & 10:00 & 11:00 \\
			95. & IU & ANMO & BHZ (20 sps) & US & 02.03.2010 & 10:00 & 11:00 \\
			: & : & : & : & : & : & : & : \\
			153. & IU & ANMO & BHZ (20 sps) & US & 29.04.2010 & 10:00 & 11:00 \\
			154. & IU & ANMO & BHZ (20 sps) & US & 30.04.2010 & 10:00 & 11:00 \\
			155. & IU & ANMO & BHZ (20 sps) & US & 01.12.2010 & 10:00 & 11:00 \\
			156. & IU & ANMO & BHZ (20 sps) & US & 02.12.2010 & 10:00 & 11:00 \\
			: & : & : & : & : & : & : & : \\
			184. & IU & ANMO & BHZ (20 sps) & US & 30.12.2010 & 10:00 & 11:00 \\
			185. & IU & ANMO & BHZ (20 sps) & US & 31.12.2010 & 10:00 & 11:00 \\ \hline
	\end{tabular}}
\end{table}

Vertical channel event-free data is acquired from the Incorporated Research Institute for Seismology (IRIS) to carry out the systematic characterization of noise. A total of $185$ datasets are carefully selected to study two crucial aspects, (i) daily variations in noise properties and model structure (case study $1$) and (ii) seasonal variations (case study $2$) in the noise properties and model structure. In both the case-studies, datasets from a fixed geographical location (ANMO station, located in the US) are considered in two different time-slots (morning 10:00 to 11:00 and night 22:00 to 23:00) during the months of March, April, and December in the year 2010. The reason for selecting these time-slots is to study the effect of cultural noise on the noise properties, and hence the model structure, during working and silent hours. Details of all the datasets are summarized in Table \ref{tab1:spec}. For analysis purposes, we have considered $50000$ data samples and subsequently, analyze the properties and model structure of seismic noise across datasets in each case study. A significance level of $0.05$ is selected for different statistical tests in all the case studies.

\subsection{Case Study $1$: Daily variations in noise properties at a given station}\label{subsec:casestudy1} 
Seismic data from ANMO station (located in New Mexico, US) for the entire duration of March – April 2010 are analyzed to study the daily variations in noise properties and model structure. Table \ref{tab1:spec} summarizes the details of all the datasets (night time: dataset 2-62; day time: dataset 94-154). The proposed method is illustrated in detail with the help of dataset $1$ with results being summarized in Figure \ref{Fig:sumamry} for the rest of the datasets.
\begin{figure}
	\vspace{-2em}
	\begin{minipage}{0.33\linewidth}
		\centering
		\includegraphics[width=\linewidth,height=0.25\textheight]{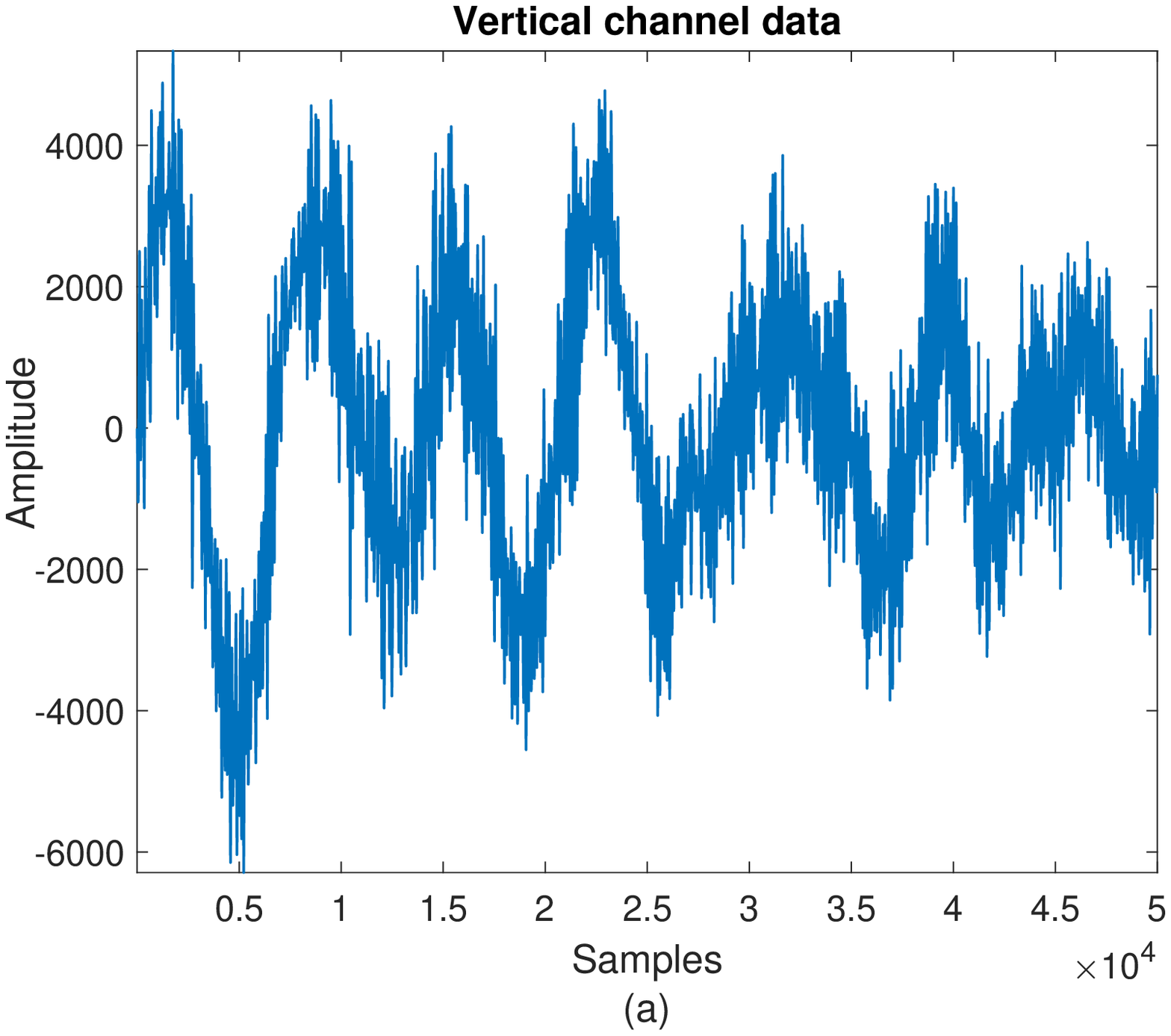}\\
		\includegraphics[width=\linewidth,height=0.25\textheight]{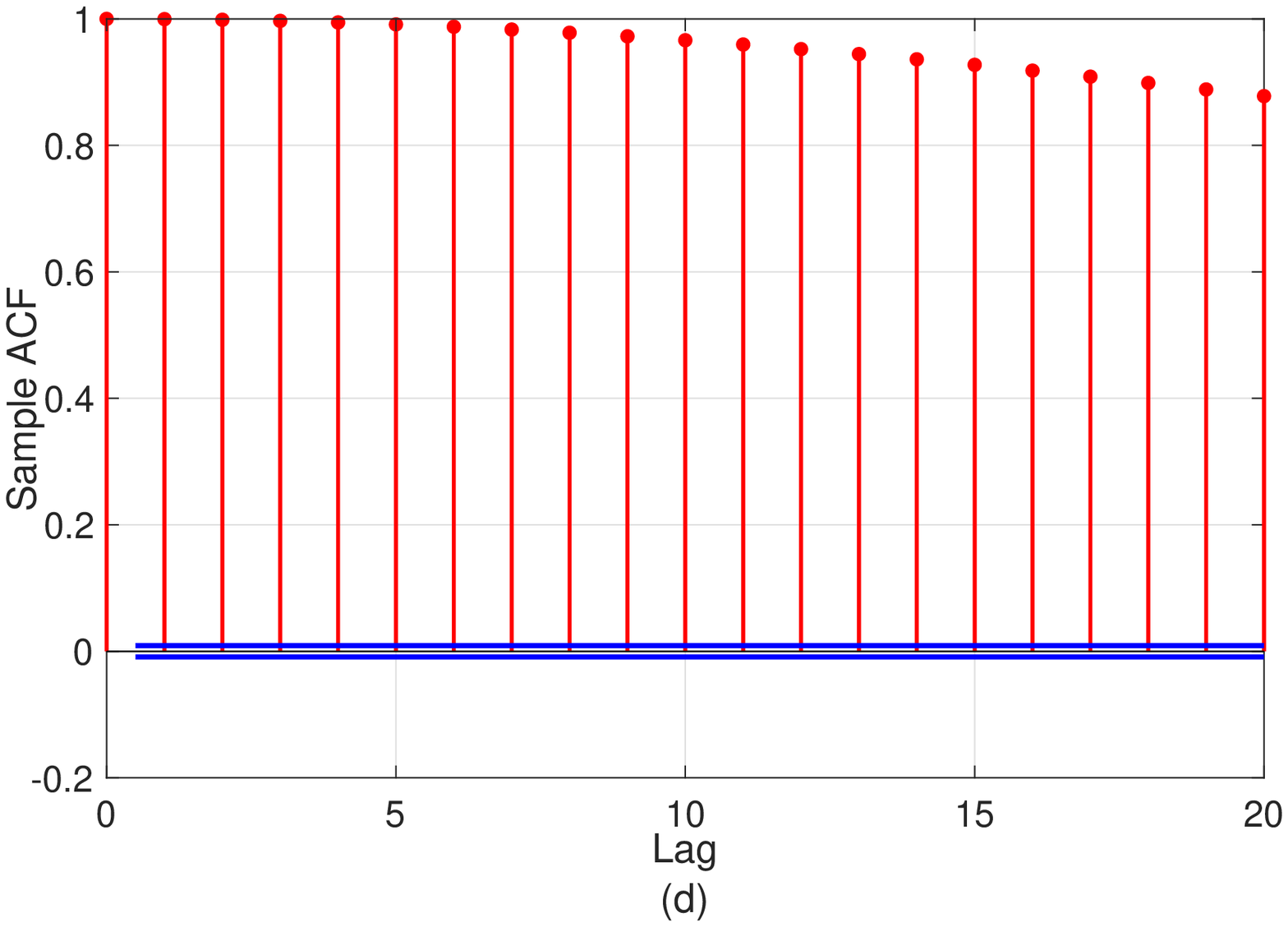}\\
		\includegraphics[width=\linewidth,height=0.25\textheight]{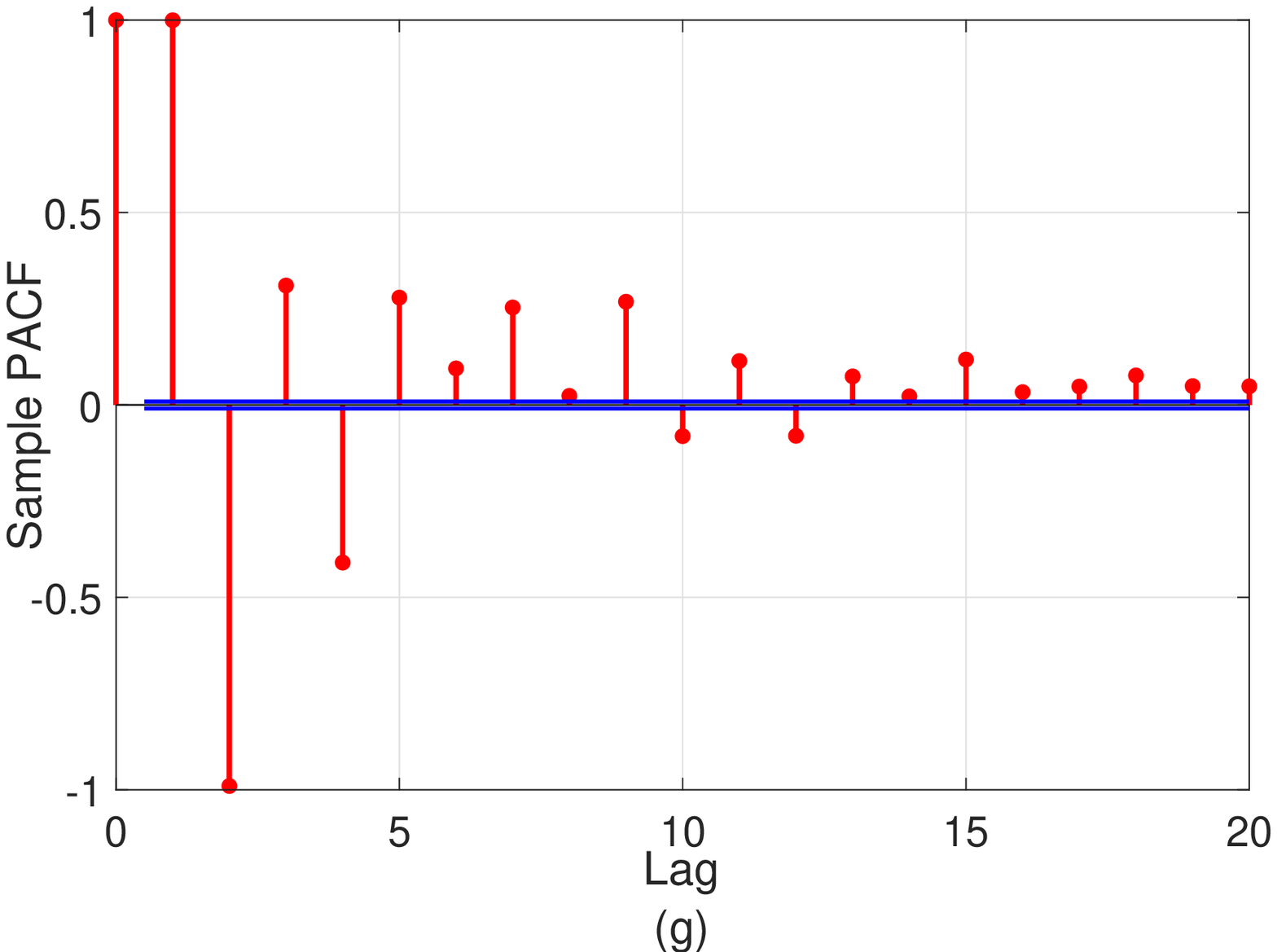}\\
		\includegraphics[width=\linewidth,height=0.25\textheight]{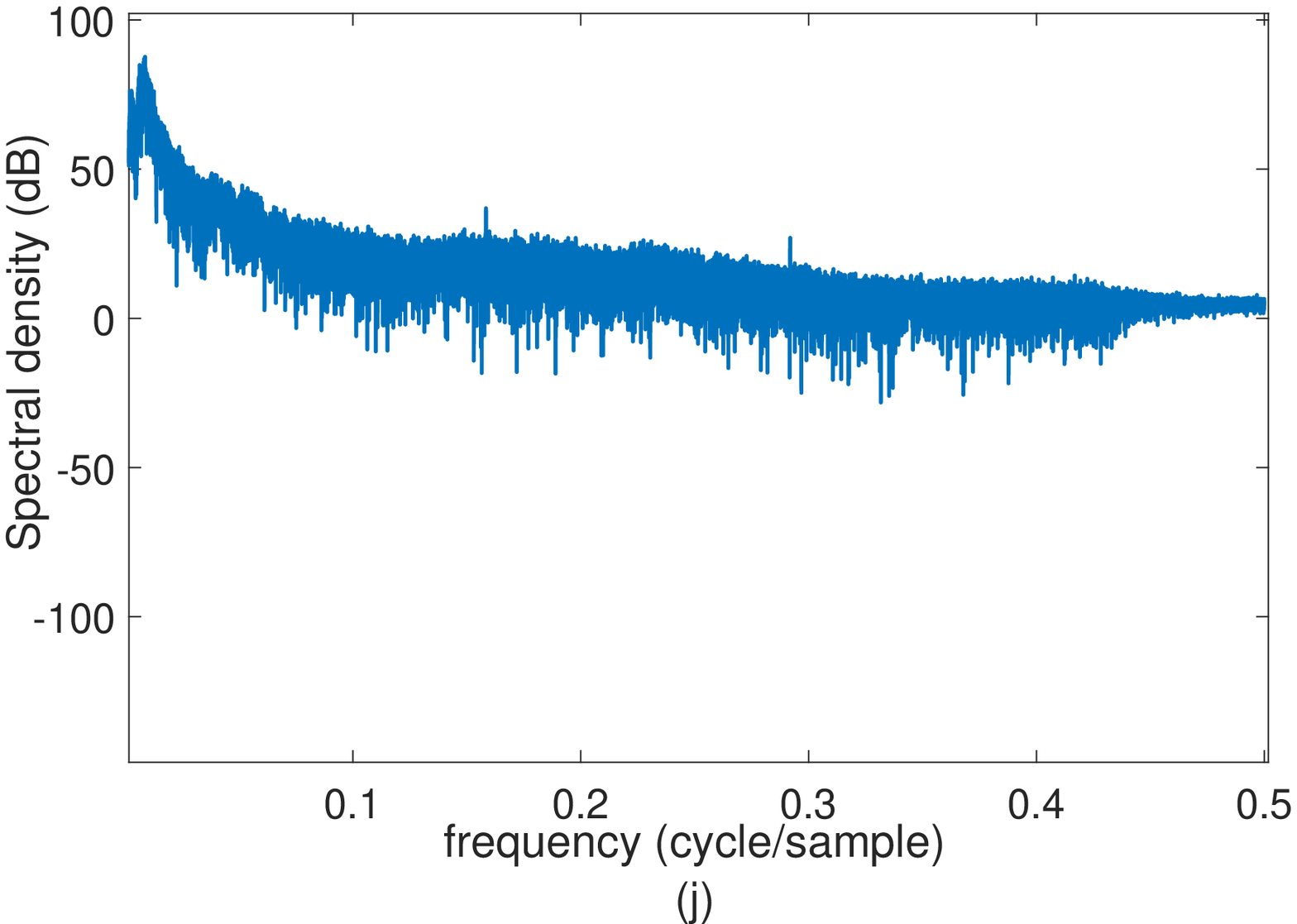}
	\end{minipage}%
	\hfill
	\begin{minipage}{0.33\linewidth}
		\centering
		\includegraphics[width=\linewidth,height=0.25\textheight]{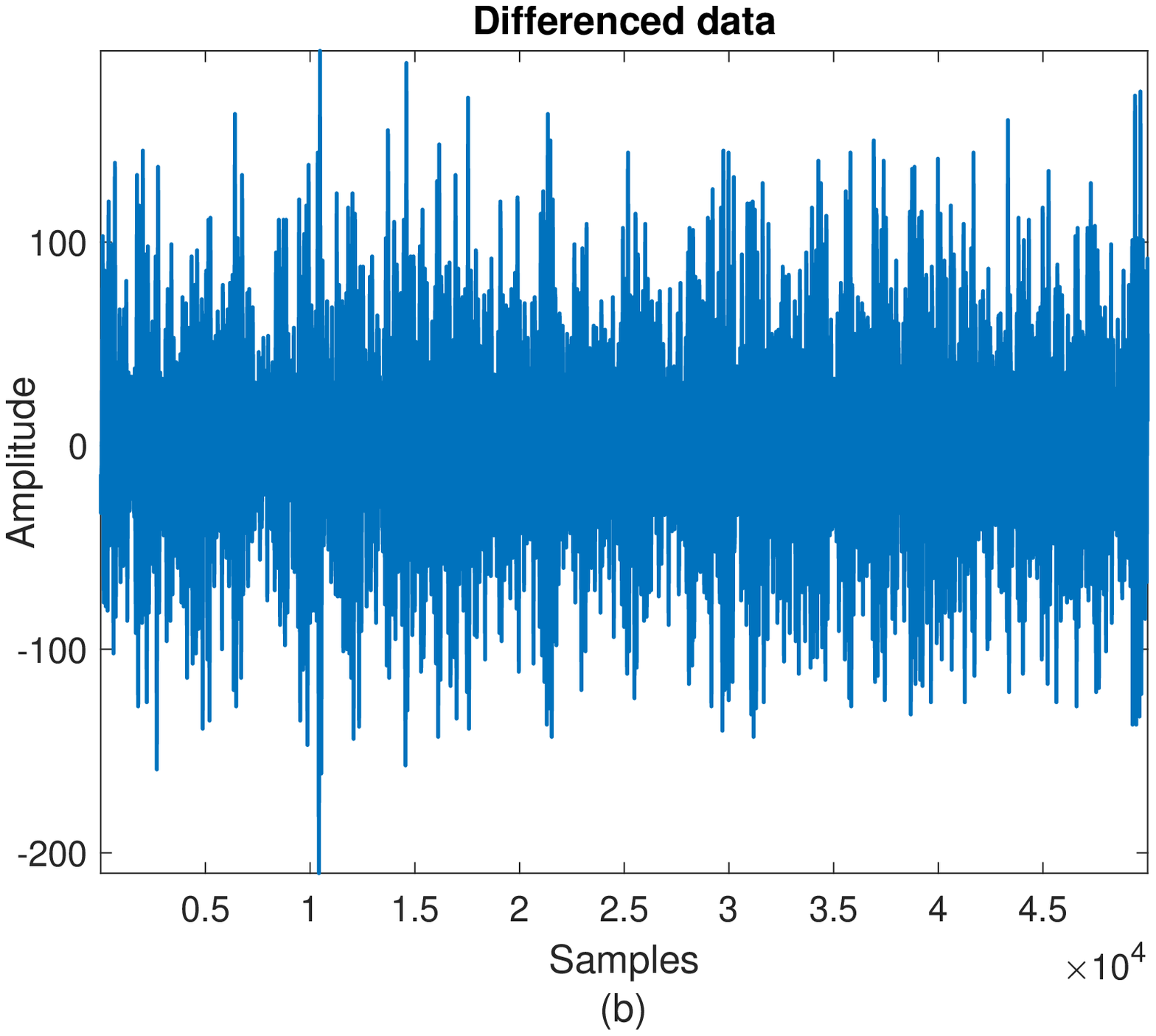}\\
		\includegraphics[width=\linewidth,height=0.25\textheight]{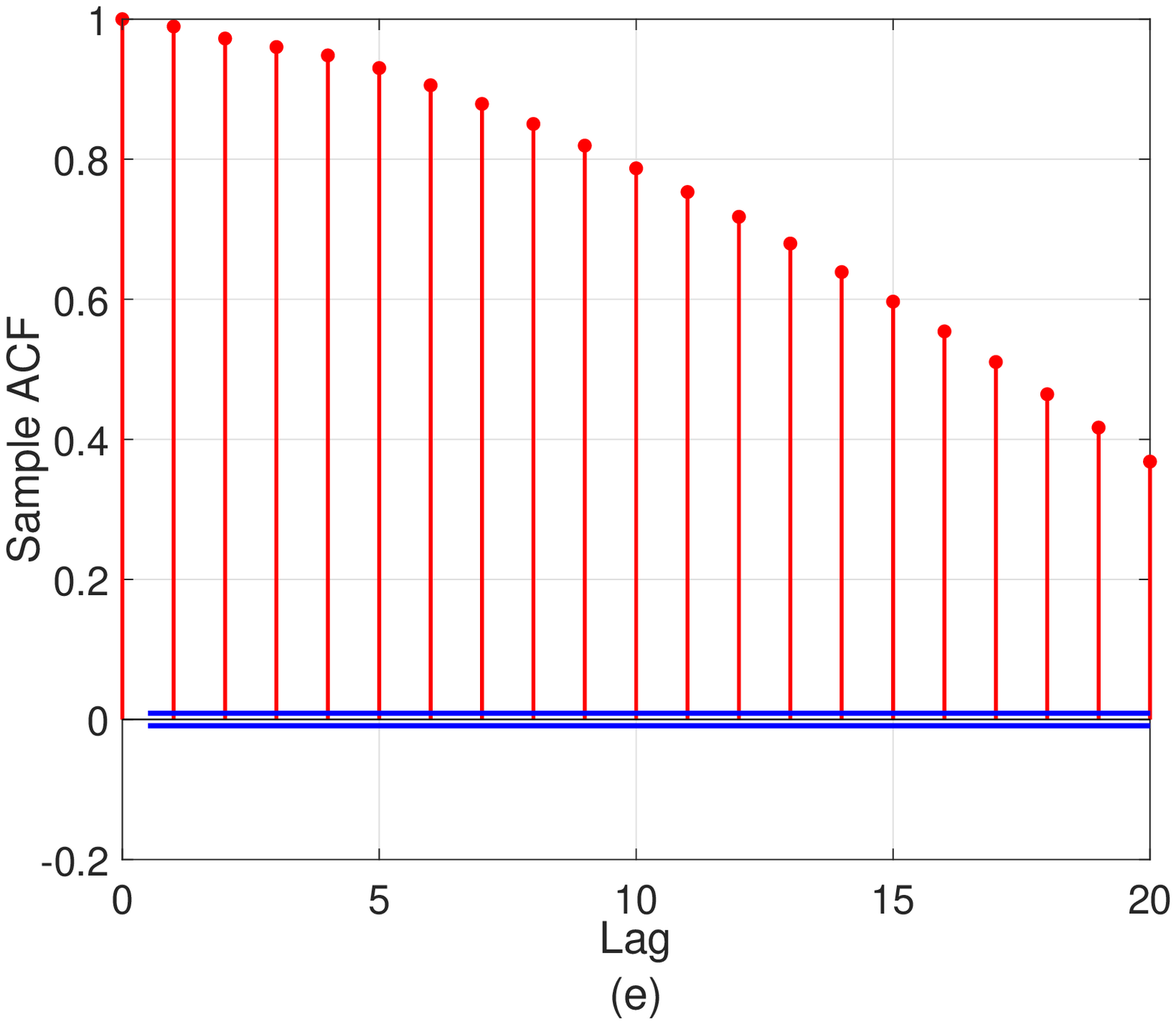}\\
		\includegraphics[width=\linewidth,height=0.25\textheight]{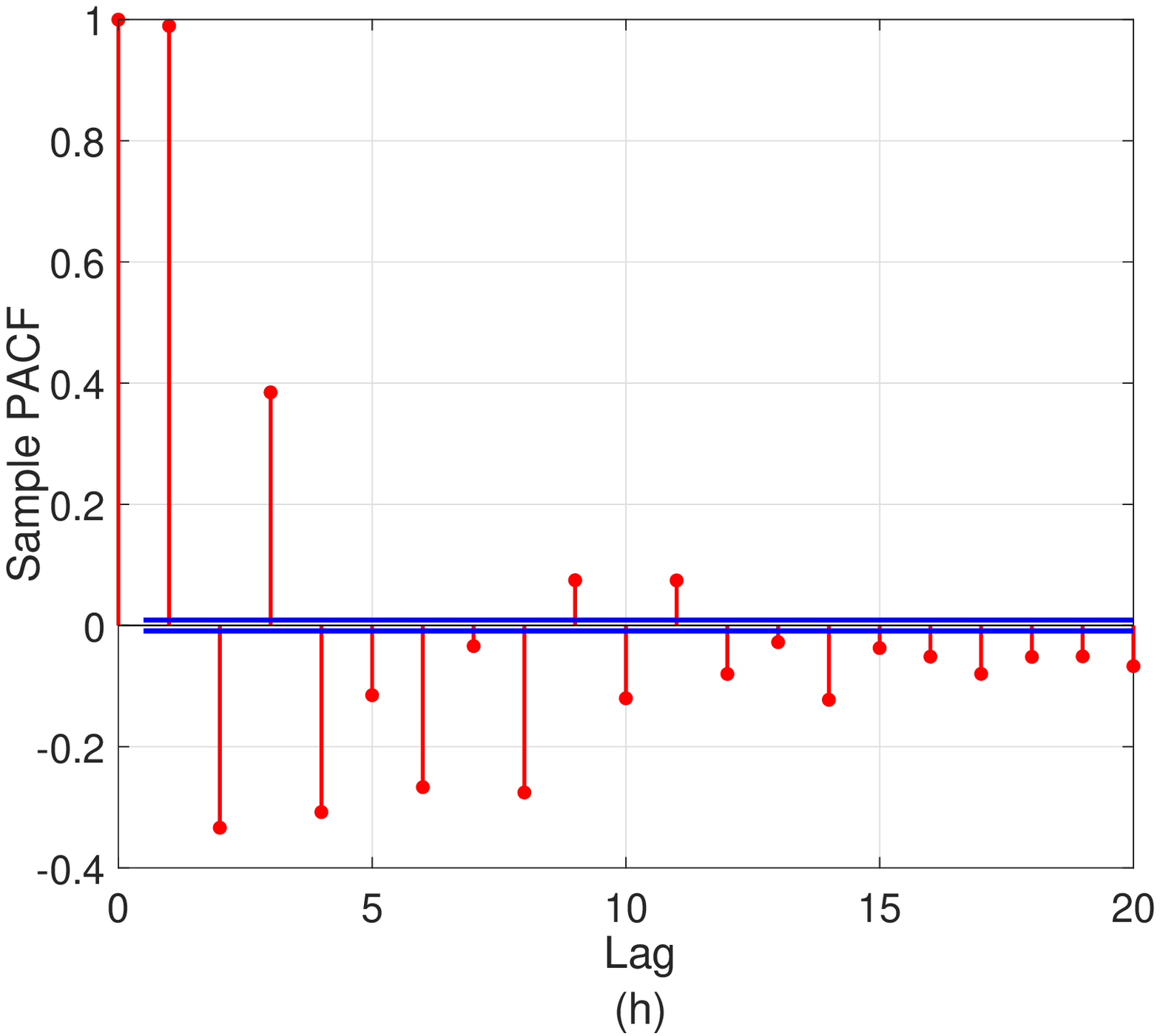}\\
		\includegraphics[width=\linewidth,height=0.25\textheight]{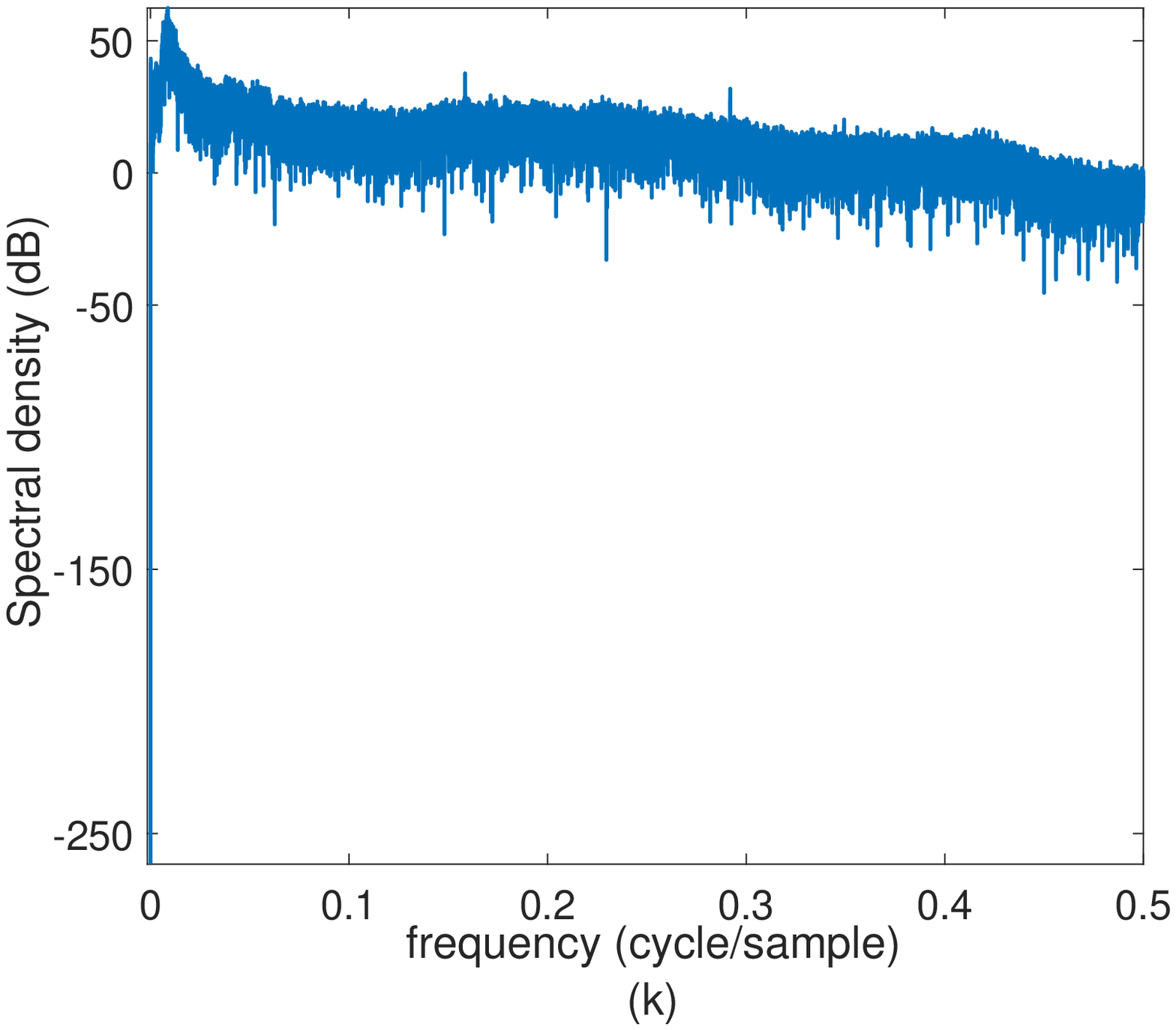}
	\end{minipage}%
	\hfill
	\begin{minipage}{0.33\linewidth}
		\centering
		\includegraphics[width=\linewidth,height=0.25\textheight]{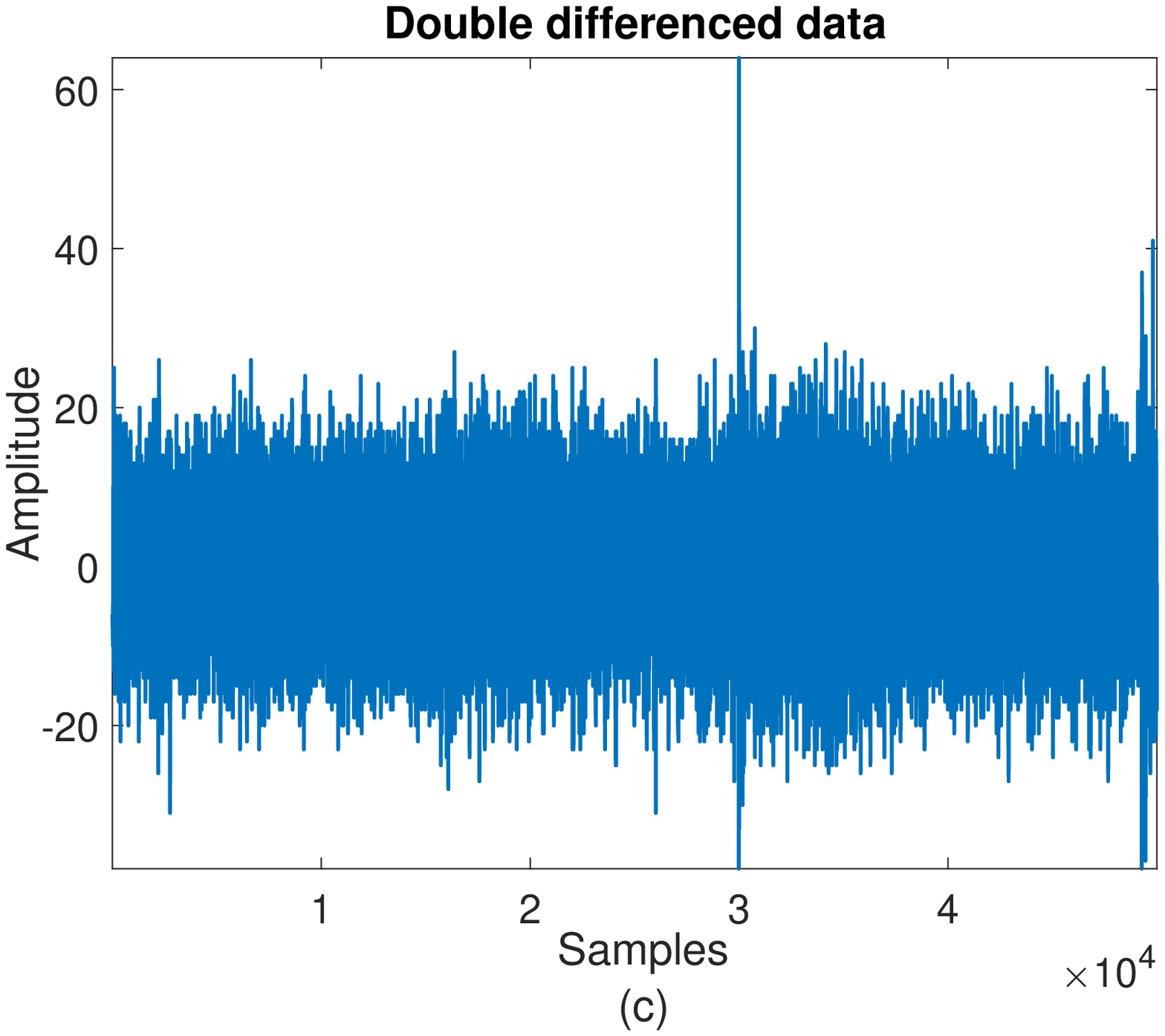}\\
		\includegraphics[width=\linewidth,height=0.25\textheight]{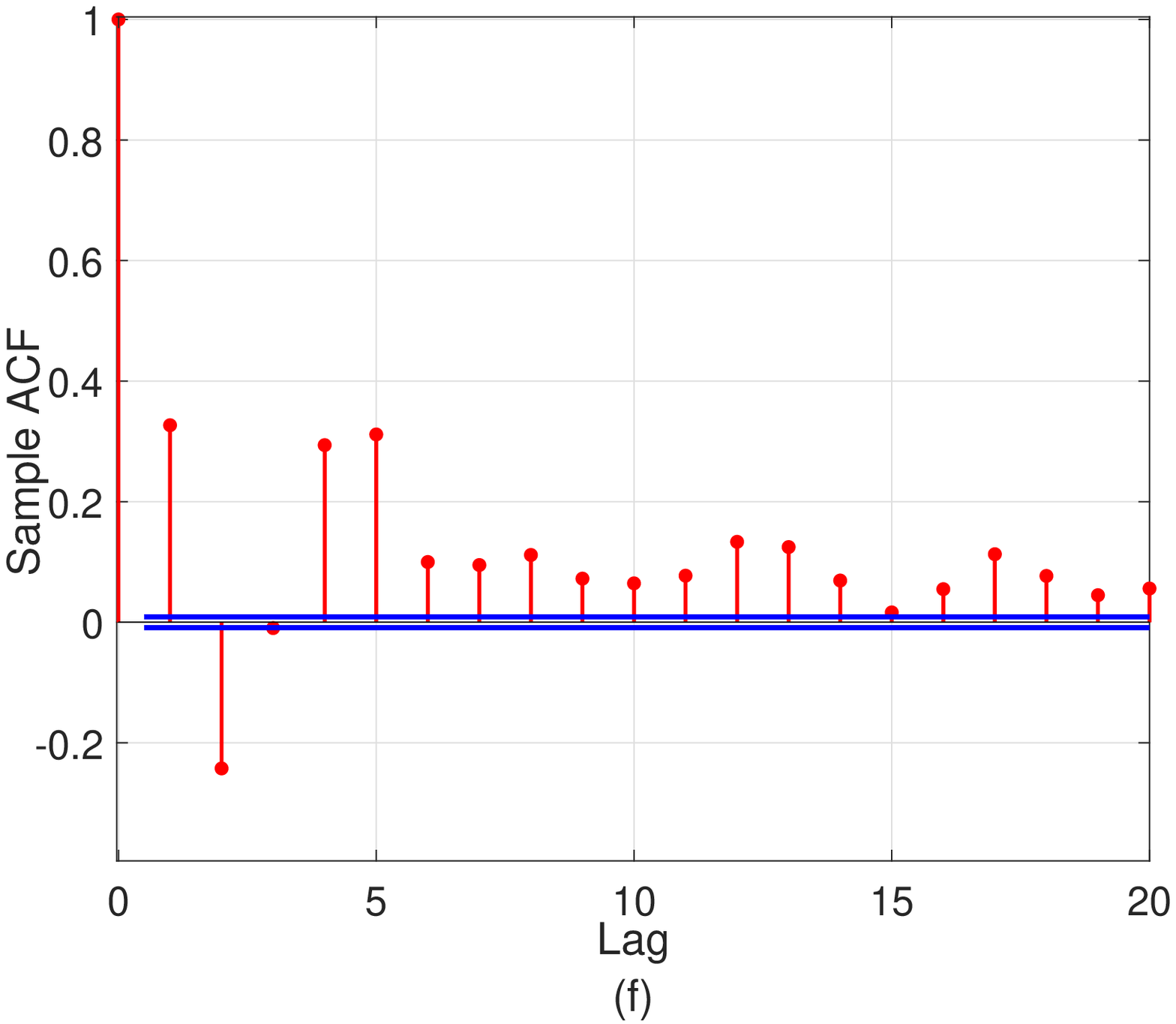}\\
		\includegraphics[width=\linewidth,height=0.25\textheight]{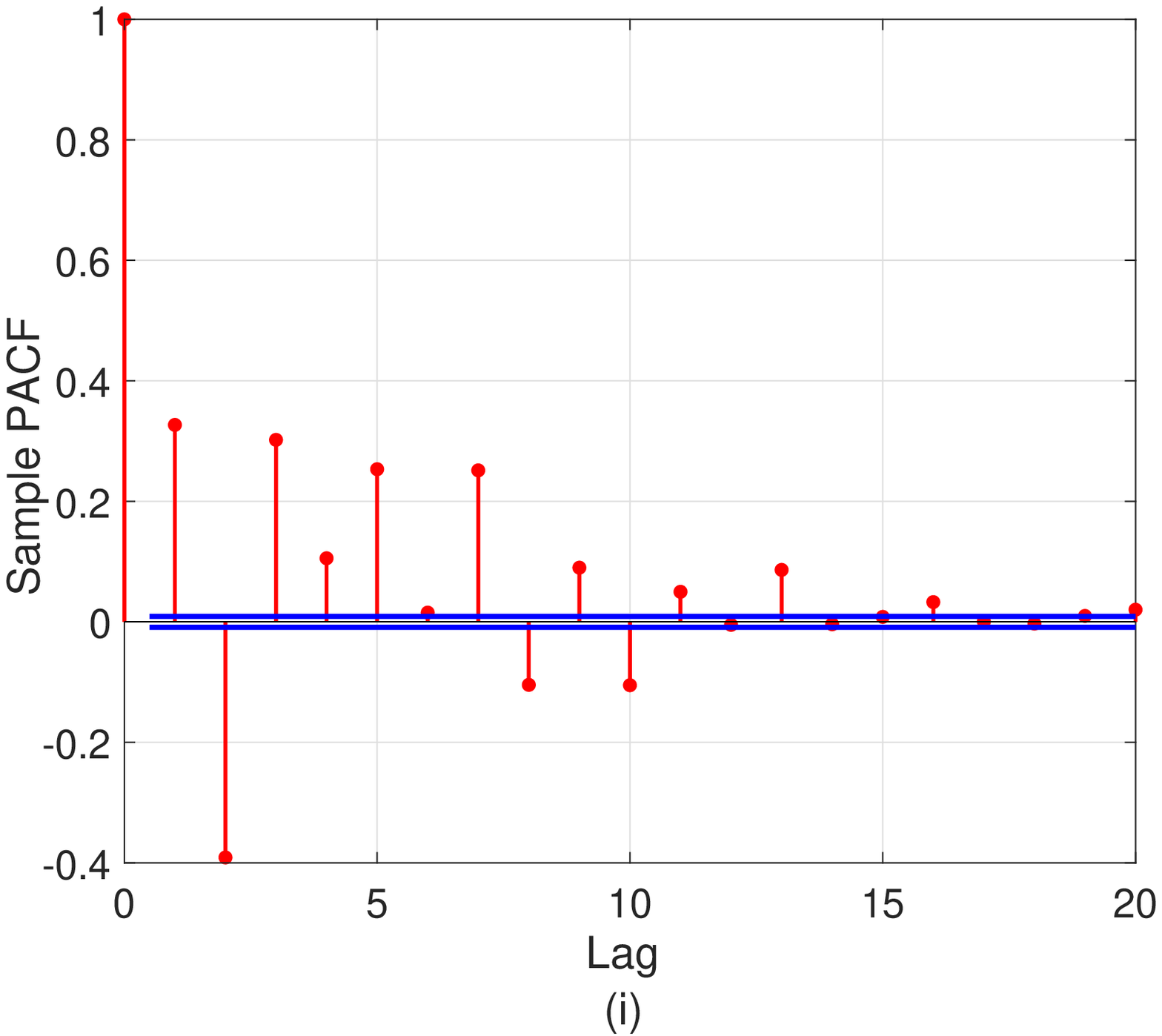}\\
		\includegraphics[width=\linewidth,height=0.25\textheight]{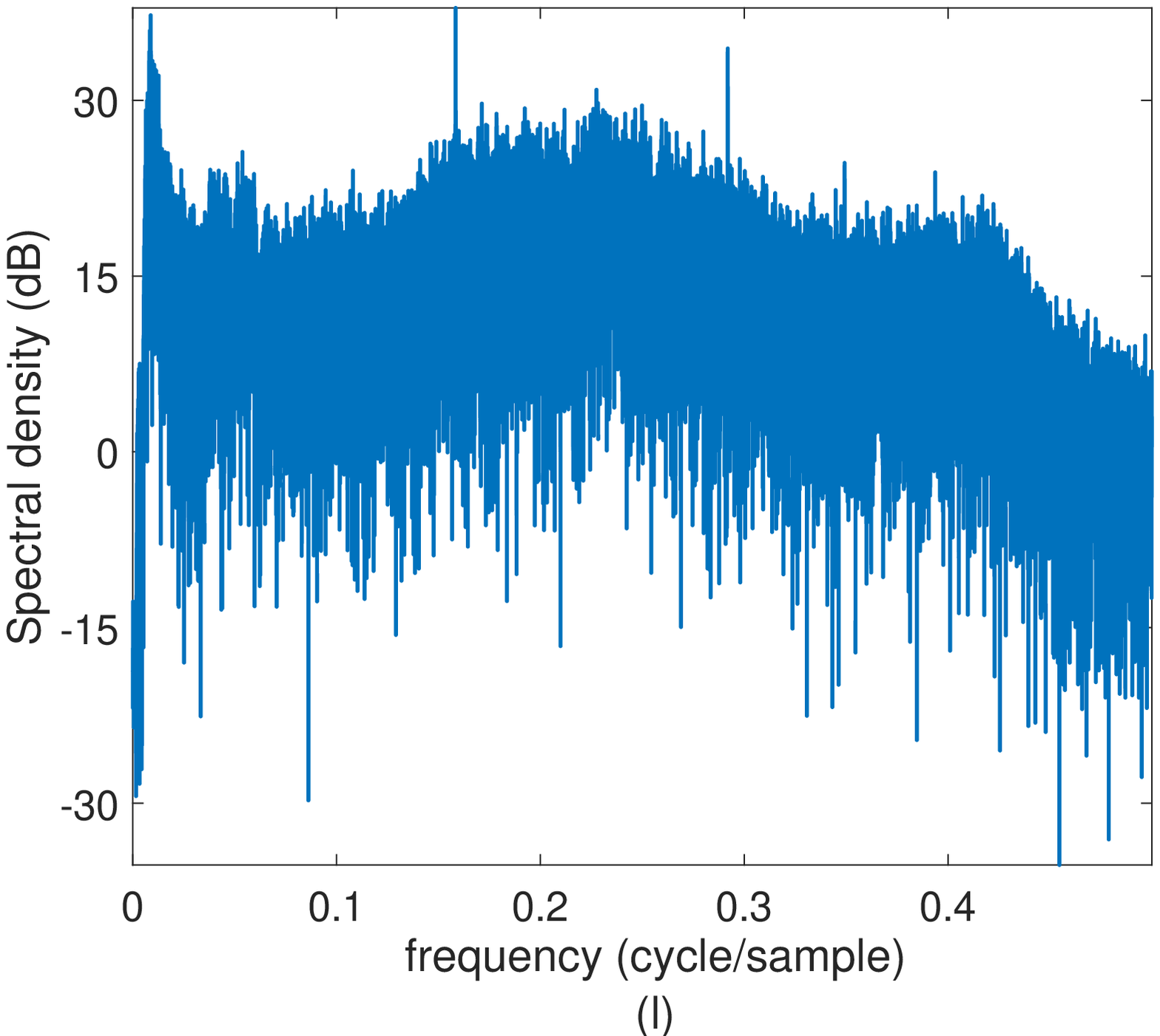}
	\end{minipage}
	\caption{Vertical channel ANMO station (a) noise, (b) differenced and (c) double differenced noise, (d,e,f) represents the ACF of noise, differenced and double differenced noise respectively, (g,h,i) are the PACF plots and (j,k,l) represents the spectral density plots for noise, differenced and double differenced noise.\label{Fig2:dataCaseStudy1}}
\end{figure}

Vertical channel seismic noise from ANMO station along with ACF, PACF and spectral density plots of the data are shown in Figure \ref{Fig2:dataCaseStudy1} (a), Figure \ref{Fig2:dataCaseStudy1} (d), Figure \ref{Fig2:dataCaseStudy1} (g), and Figure \ref{Fig2:dataCaseStudy1} (j) respectively. Slowly decaying nature of ACF and near-unity value of PACF at lag $1$ and $2$ indicates the presence of integrating effect of order $2$ in the ANMO station data. This qualitative inference is supported statistically using ADF and PP tests on the data whose results are summarized in Table \ref{Tab:stat_test}. The observed value for both ADF and PP tests is greater than the critical value for the raw data, while the observed value is lesser than the critical value for differenced data. It confirms the presence of unit root of order $1$. Therefore, the data is differenced once to model the integrating effects. Differenced data is then tested for the presence of heteroskedasticity using the PSR test. The results for the PSR test for raw and differenced data are reported in Table \ref{Tab:stat_test}. PSR test reveals the presence of heteroskedasticity in data. Both ADF and PP tests result in false rejection of the null hypothesis in the presence of heteroskedastic errors. We, therefore, analyze the estimated AR($1$) model coefficients to verify the results obtained from the ADF and PP tests. Estimated AR($1$) coefficients for raw, differenced, and double differenced data ($1$, $0.99$, and $0.48$ respectively) reveal that the correct order of integrating effects is $2$. Therefore, differencing of degree $2$ is required to make the data stationary. First-order stationary noise is then tested for linearity using a surrogate-based method. In this work, twenty surrogates are generated using amplitude adjusted Fourier transform (AAFT) to conduct the linearity test. Surrogate-based linearity test (Table \ref{Tab:stat_test}) reveals the non-linear nature of ANMO noise. It is also observed from the linearity test that the differencing operation does not hamper the linearity / non-linearity in data.

\begin{figure}
	\centering
	\includegraphics[scale=0.35]{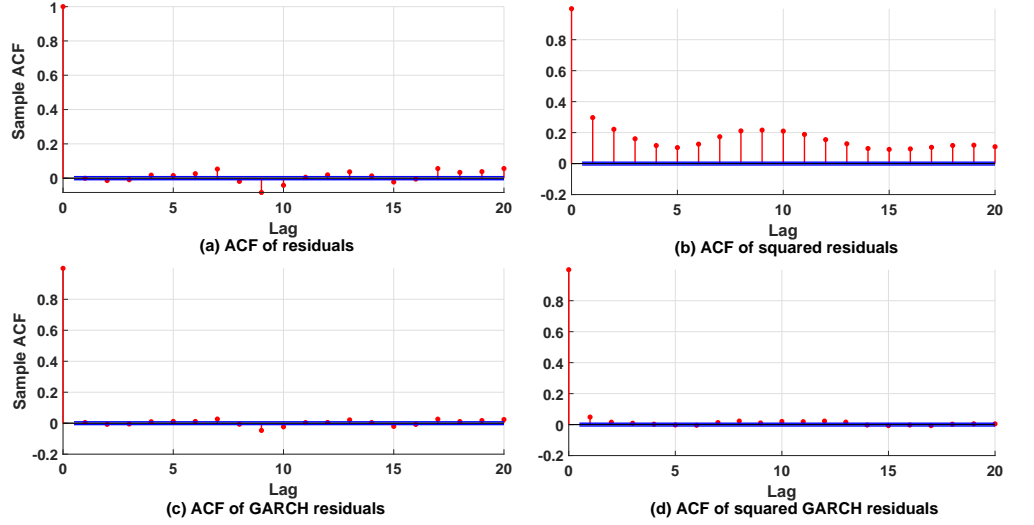}
	\caption{ACF of (a) ARIMA residuals, (b) squared ARIMA residuals, (c) GARCH residuals and (d) squared GARCH residuals\label{Fig:ANMO_res}}
\end{figure}

An ARIMA model of suitable order is developed for the nonlinear ANMO station noise because the scope of this work is limited to linear models only. We know violation of linearity will result in sub-optimal predictions, hence, we do not expect the residuals to be white. Residual and error analysis of a range of estimated models suggest an ARIMA($5,2,3$) model best fits the ANMO station noise. Estimated model along with the standard errors is given by \eqref{eq:modeq}.

\begin{subequations}
	\label{eq:modeq}
	\begin{align}
	y[k] & = \dfrac{C(q^{-1})}{(1-q^{-1})^{2}D(q^{-1})} e[k] \\
	C(q^{-1}) & = 1 - \underset{(\pm 0.008)}{0.54 }q^{-1} - \underset{(\pm 0.007)}{0.62}q^{-2} +  \underset{(\pm 0.006)}{0.63}q^{-3} \\
	D(q^{-1})  = 1 - \underset{ (\pm 0.009)}{1.29}q^{-1} + &\underset{ (\pm 0.013)}{0.39}q^{-2} + \underset{ (\pm 0.012)}{0.17}q^{-3} - \underset{ (\pm 0.01)}{0.29}q^{-4} + \underset{ (\pm 0.006)}{0.24}q^{-5}  
	\end{align}
\end{subequations}	
\noindent where, $e[k]\sim \mathcal{N}(0,44.91)$. All the estimated parameters are significant as compared to their standard errors implying that the model is not under-fitted. Moreover, as observed from Figure \ref{Fig:ANMO_res}, residuals of estimated ARIMA model are completely white implying that the estimated ARIMA model for dataset $1$ is satisfactory.

\begin{table}
	\centering
	\caption{Summary of statistical tests for ANMO station seismic noise where ADF $\&$ PP are left-tail (LT) test, linearity test is two-tail (TT) test and PSR $\&$ ARCH are right-tail (RT) test. RNH stands for reject null hypothesis.\label{Tab:stat_test}}
	\centering
	\resizebox{\linewidth}{!}{
		\begin{tabular}{@{}llcccccl@{}}
			\hline
			\textbf{Test} & \multicolumn{1}{c}{\textbf{H$_{0}$}} & \textbf{$p$-value} & \textbf{\begin{tabular}[c]{@{}c@{}}Observed\\ value\end{tabular}} & \textbf{\begin{tabular}[c]{@{}c@{}}Lower\\ critical\\ value\end{tabular}} & \textbf{\begin{tabular}[c]{@{}c@{}}upper\\ critical\\ value\end{tabular}} & \textbf{\begin{tabular}[c]{@{}c@{}}Nature of\\ test\end{tabular}} & \textbf{Outcome} \\ \hline
			\multicolumn{8}{c}{\textbf{Tests on data}} \\ \hline
			\textbf{ADF test} & \begin{tabular}[c]{@{}l@{}}Unit root\\ is present\end{tabular} & 0.24 & -2.70 & -3.41 & - & LT & \begin{tabular}[c]{@{}l@{}}Cannot RNH\\ Unit root\end{tabular} \\
			\textbf{PP test} & Integrating effect & 0.24 & -2.70 & -3.41 & - & LT & \begin{tabular}[c]{@{}l@{}}Cannot RNH\\ Integrating effect\end{tabular} \\
			\textbf{\begin{tabular}[c]{@{}l@{}}PSR test\\ T\\ I+R\\ T+I+R\end{tabular}} & \begin{tabular}[c]{@{}l@{}}Series is\\ homoskedastic\end{tabular} & \begin{tabular}[c]{@{}l@{}}0\\ 0\\ 0\end{tabular} & \begin{tabular}[c]{@{}l@{}}339.07\\ 6041.49\\ 6380.57\end{tabular} & - & \begin{tabular}[c]{@{}l@{}}23.68\\ 4009.72\\ 4023.98\end{tabular} & RT & \begin{tabular}[c]{@{}l@{}}RNH\\ Heteroskedastic\end{tabular} \\ 
			\textbf{Linearity test} & Series is linear & - & 3.32 & 3.93 & 4.02 & TT & \begin{tabular}[c]{@{}l@{}}RNH\\ Non-linear\end{tabular} \\ \hline
			\multicolumn{8}{c}{\textbf{Tests on differenced data}} \\ \hline
			\textbf{ADF test} & \begin{tabular}[c]{@{}l@{}}Unit root\\ is present\end{tabular} & 0.001 & -28.19 & -1.94 & - & LT & \begin{tabular}[c]{@{}l@{}}RNH no\\ unit root\end{tabular} \\
			\textbf{PP test} & Integrating effect & 0.001 & -27.45 & -1.94 & - & LT & \begin{tabular}[c]{@{}l@{}}RNH no\\ integrating effect\end{tabular} \\
			\textbf{\begin{tabular}[c]{@{}l@{}}PSR test\\ T\\ I+R\\ T+I+R\end{tabular}} & \begin{tabular}[c]{@{}l@{}}Series is\\ homoskedastic\end{tabular} & \begin{tabular}[c]{@{}l@{}}0\\ 0\\ 0\end{tabular} & \begin{tabular}[c]{@{}l@{}}360.22\\ 6021.15\\ 6381.38\end{tabular} & - & \begin{tabular}[c]{@{}l@{}}23.68\\ 4009.72\\ 4023.98\end{tabular} & RT & \begin{tabular}[c]{@{}l@{}}RNH\\ Heteroskedastic\end{tabular} \\
			\textbf{Linearity test} & Series is linear & - & 3.12 & 3.83 & 3.95 & TT & \begin{tabular}[c]{@{}l@{}}RNH\\ Non-linear\end{tabular} \\ 
			\hline
			\multicolumn{8}{c}{\textbf{Tests on pre-whitened data}} \\ \hline
			\textbf{SW test} & \begin{tabular}[c]{@{}l@{}}Normal\\ distribution\end{tabular} & 0.19 & 0.997 & - & - & - & \begin{tabular}[c]{@{}l@{}}cannot RNH\\ Normal distribution\end{tabular}\\
			\textbf{\begin{tabular}[c]{@{}l@{}}PSR test\\ T\\ I+R\\ T+I+R\end{tabular}} & \begin{tabular}[c]{@{}l@{}}Series is\\ homoskedastic\end{tabular} & \begin{tabular}[c]{@{}l@{}}0\\ 0\\ 0\end{tabular} & \begin{tabular}[c]{@{}l@{}}379.12\\ 6080.02\\ 6459.15\end{tabular} & - & \begin{tabular}[c]{@{}l@{}}23.68\\ 4009.72\\ 4023.98\end{tabular} & RT & \begin{tabular}[c]{@{}l@{}}RNH\\ Heteroskedastic\end{tabular} \\
			\textbf{ARCH test} & \begin{tabular}[c]{@{}l@{}}No ARCH\\ effects\end{tabular} & 0 & 15804.1 & - & 3.84 & RT & \begin{tabular}[c]{@{}l@{}}RNH\\ ARCH effect\end{tabular}\\
			\hline
	\end{tabular}}
\end{table}

Statistical tests for Gaussianity and heteroskedasticity assume that underlying data is uncorrelated or weakly correlated. However, it is observed from Figure \ref{Fig2:dataCaseStudy1}(d) and Figure \ref{Fig2:dataCaseStudy1}(g) that the seismic noise is highly correlated. Therefore, we work with the residuals obtained from ARIMA($5,2,3$) model for further characterization. It is observed from Table \ref{Tab:stat_test} that the $p$-value for SW test ($0.997$) is greater than the selected significance level ($0.05$). A higher $p$-value indicates that the test fails to reject the null hypothesis that the residuals (and hence the ANMO station noise) are driven by a Gaussian process. For testing heteroskedasticity, it is observed from Figure \ref{Fig:ANMO_res}(a) and Figure \ref{Fig:ANMO_res}(b) that the ACF of residuals exhibits white noise characteristics while the ACF of squared residuals is significant. Therefore, it is safe to say that the residuals exhibits conditional heteroskedasticity. This qualitative inference is also verified quantitatively using PSR and ARCH test. The results of these statistical tests (Table \ref{Tab:stat_test}) also support the inference drawn from the visual inspection of squared residuals. It is also observed from the spectral density plot of data (Figure \ref{Fig2:dataCaseStudy1}(j)) that the seismic noise contains the effect of microseisms at the expected frequencies. Furthermore, the differencing operation reduces the contribution of lower frequencies to the total energy of the seismic noise.

The conditional heteroskedasticity in residuals is modeled using GARCH models of suitable order. Residual and error analysis of a range of estimated GARCH models suggest GARCH($1,1$) to be the most suitable model. The estimated GARCH model for the residuals obtained from  ARIMA($5,2,3$) is given by equation \eqref{eq:ANMO_garch}.

\begin{eqnarray}
~~~~~~&e[k] = \sigma_{k}z[k], ~~~~z[k] \sim \mathcal{N}(0,1) \nonumber\\
\sigma^{2}_{k} &= \underset{(\pm 0.03)}{0.31} + \underset{(\pm 0.001)}{0.98} \sigma^{2}_{k-1} +  \underset{(\pm 0.0009)}{0.019} e^{2}_{k-1} 
\label{eq:ANMO_garch}
\end{eqnarray}

Both the residuals and squared residuals of the estimated GARCH model exhibit white noise characteristics as indicated by Figure \ref{Fig:ANMO_res}(c,d). In this case, non-linearity in the noise is modeled by the ARIMA$-$GARCH model. However, this is not always the case. It can, therefore, be said that seismic noise at ANMO station tested positive for two specific types of non-stationarities, (i) integrating effects of order $2$ and (ii) heteroskedasticity, Gaussianity and conditional heteroskedasticity. With respect to linearity, ANMO station noise tested negative. Moreover, the noise can be modeled by ARIMA($5,2,3$)-GARCH($1,1$) model.

The same systematic procedure is applied to all the datasets to study daily variation in the properties and model structure. As observed from Figure \ref{Fig:sumamry} (spring season), $114$ out of $122$ datasets tested positive for the presence of integrating effects of order $2$, while the remaining datasets exhibit first-order integrating effects. However, only $24$ datasets tested negative for linearity. All the datasets tested positive for normality, heteroskedasticity, and conditional heteroskedasticity. It is also observed that the characteristics such as non-stationarities, the order of integrating effects, and Gaussianity do not vary with day or night time (Figure \ref{Fig:sumamry} day and night time data), unlike the linearity property. A speculated reason for such an observation is that the impact of cultural noise is much less during night time than in the day time, resulting in the changing properties over time. Furthermore, features that do not change can be attributed to the instrumentation part or the nature of the generating process. Since, the properties of seismic noise are not changing on daily basis (except linearity), model structure for ANMO station can be fixed.

\begin{figure}
	\centering
	\includegraphics[scale=0.35]{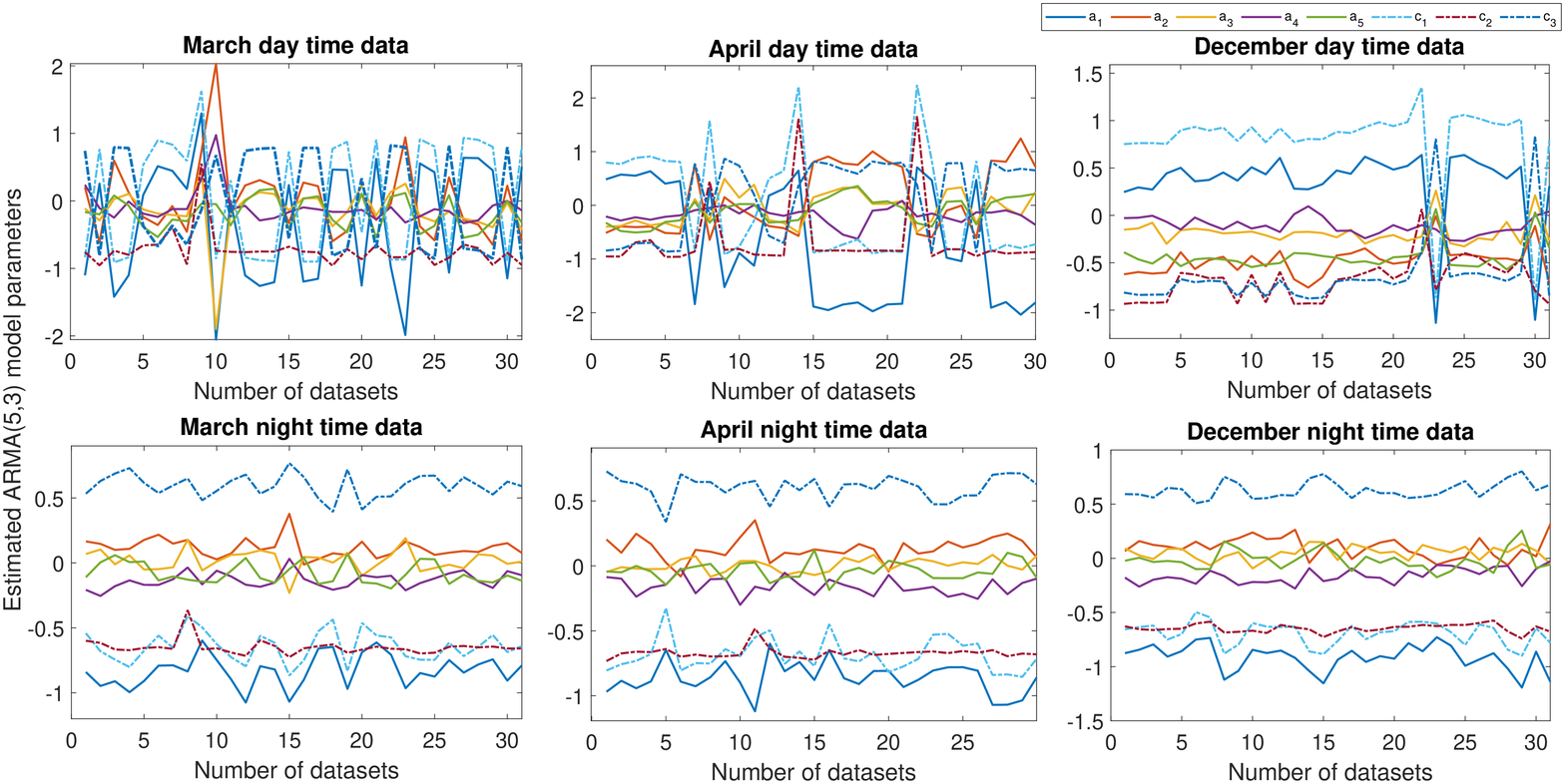}
	\caption{Variation in estimated ARIMA(5,3) model parameters for all the datasets, where the dashed line represents the MA parameters and solid lines represents the AR component of the model. The parameters are estimated on the double differenced data.\label{Fig:param_var}}
\end{figure}

\begin{figure}
	\centering
	\includegraphics[scale=0.45]{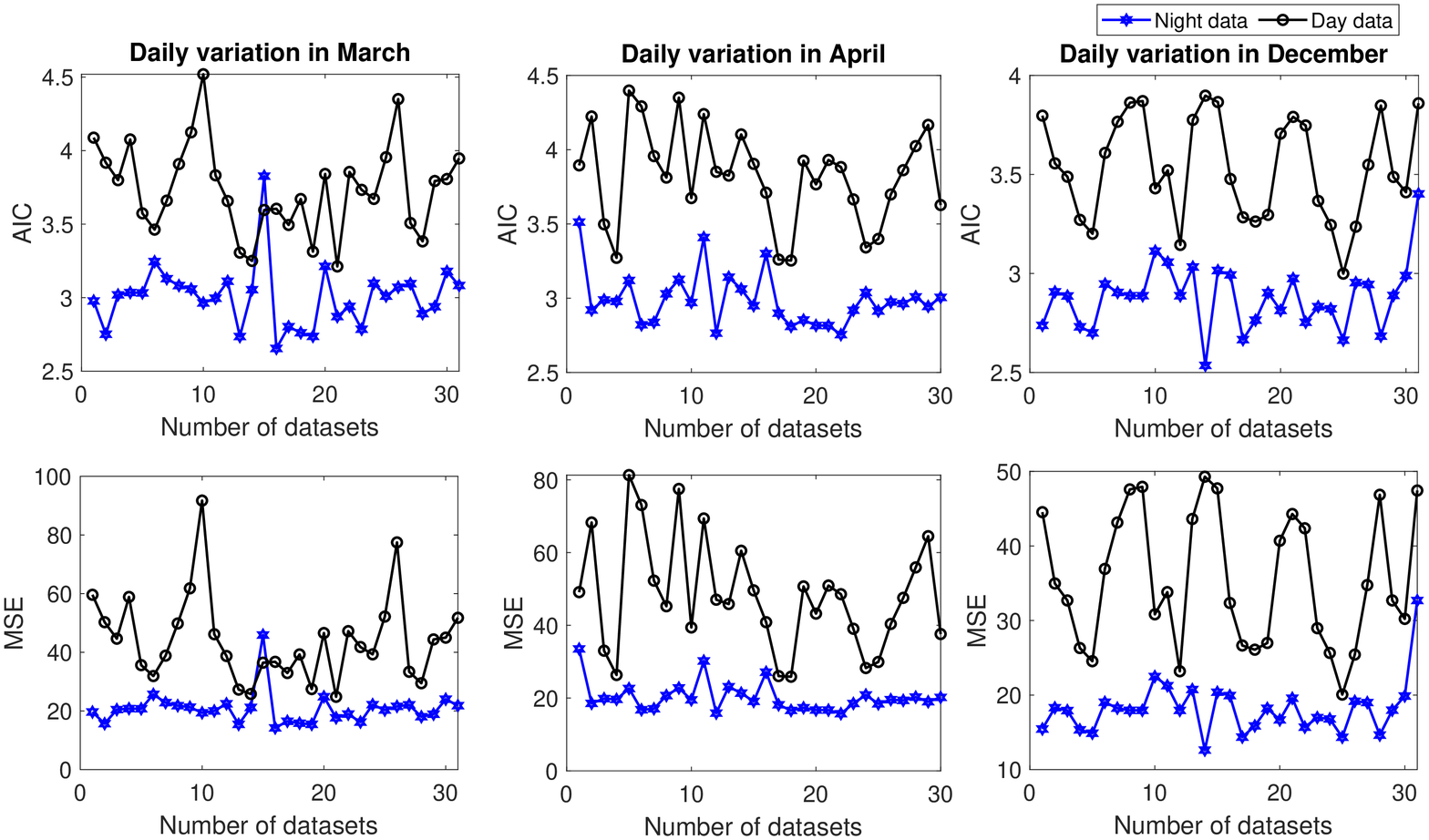}
	\caption{Variation in AIC and mean square error (MSE) obtained from ARIMA(5,2,3) model for all the datasets.\label{Fig:AIC_var}}
\end{figure}

\begin{figure}
	\centering
	\includegraphics[scale=0.45]{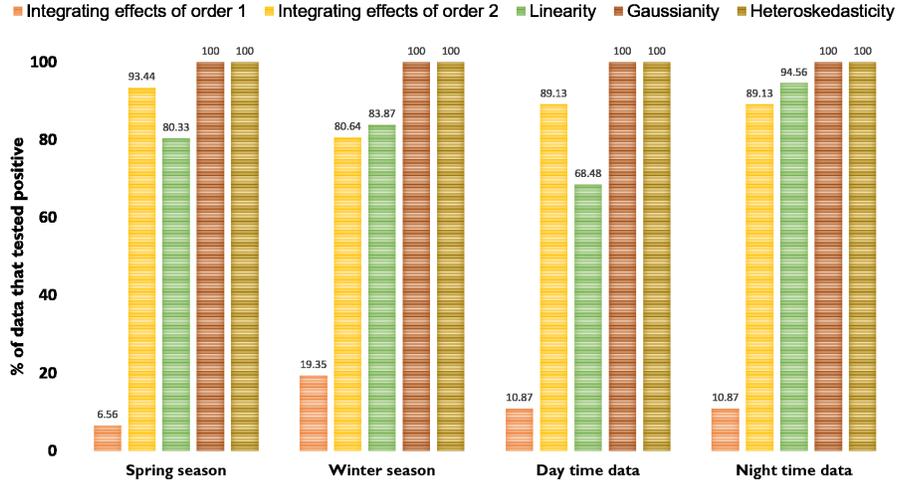}
	\caption{Visualization of test results for variations in properties with spring season ($122$ datasets), winter season ($62$ datasets), day and night time data ($92$ datasets each). Each color bar represents a specific property and the numbers on top of the bar represents the percentage of datasets that tested positive for that particular property.\label{Fig:sumamry}}
\end{figure}

For ANMO station, ARIMA($5,2,3$) results in a satisfactory model. However, as observed from Figure \ref{Fig:param_var} and Figure \ref{Fig:AIC_var}, the estimated model parameters vary with the time slots. It is also observed that the variation in estimated parameters is higher during the day time compared to the night time (silent hours). Moreover, the variation in estimated parameters increases during the festive time of Christmas and New year. As observed from day time data in the month of December Figure \ref{Fig:param_var}, there is a sudden change in the model parameters on the Christmas and New year eves for day time data. This sudden change in the estimated parameters is due to the increased human-induced cultural noise during the festive time. It is also observed that the estimated ARIMA (5,2,3) parameters are constant during night time irrespective of the festivity. Therefore, it can be said that for a fixed geographical location, noise properties, and hence, the model structure can be fixed irrespective of the time. However, due to the variation in human-induced cultural noise during day and night time, the model needs to be updated every day. Since the contribution of cultural noise is minimal during the night time, the estimated model can also be fixed for the night time data.

These significant discoveries of the work, namely, the presence of integrating effects and heteroskedasticity in seismic noise pertain to the data acquisition and the physics of generating process. Attributing each of these statistical properties to a particular source requires a detailed analysis and is restricted for future studies. At present, we speculate that the presence of integrating effects are due to the instrumentation while the heteroskedasticity is due to the change in properties of the seismic wave as it travels through different layers. The non-linearity in noise, in retrospect, is not so surprising because of the heterogeneous medium the seismic wave travels through. Estimated time-series model, which is commensurate with the noise properties, is beneficial for various applications such as model-based techniques for detecting P-wave onset in seismic signals.

\subsection{Case Study $2$: Seasonal variations in noise properties at a given station }\label{subsec:casestudy3}
Data for two different seasons, spring (March-April) and winter (December), with the same time slots as considered in case study $1$ is analyzed. Details of all the datasets are summarized in Table \ref{tab1:spec} (day time: datasets 2-62 for spring $\&$ datasets 63-93 for winter; night time: 94-154 for spring $\&$ 155-185 for winter season). The proposed systematic procedure is applied to all the $184$ datasets ($122$ for spring and $62$ for winter season). During the winter season, it is observed that $50$ datasets out of $62$ exhibit integrating type non-stationarity of order $2$, and the rest tested positive for the integrating effects of order $1$.  The datasets collected in the month of December also tested positive for heteroskedasticity, conditional heteroskedasticity, and are driven by Gaussian white noise. For linearity test, $52$ out of $62$ tested positive for linearity. It is observed by comparing the results across seasons (results for the spring season are discussed in the case study $1$) that there is no specific change in the properties with seasons.  Figure \ref{Fig:sumamry} depicts that noise properties do not vary with day and night time irrespective of the seasonal variation, except for the linearity feature. Therefore, it can be said that climatic conditions or seasonal variations do not alter the noise properties. Hence, the model structure can also be fixed for seasonal variations. It is a known fact in seismic literature that the noise is affected by climatic changes (\cite{wilcock1999effect,peterson1993observations}), which is also observed in our analysis. As observed from Figure \ref{Fig:param_var}, variation in the estimated model parameters are lesser during the winter season as compared to the spring season. For the winter season, estimated mean square error (MSE) and Akaike-Information Criterion (AIC) shown in Figure \ref{Fig:AIC_var} are much smaller as compared to the spring season.

\section{Conclusions}\label{sec:Conclusions}This work proposed a systematic methodology to characterize the seismic noise and develop a time-series model that is commensurate with the noise properties. With respect to our findings, two novel statistical features of seismic noise, namely, the integrating effect and heteroskedasticity, were discovered. The latter feature is particularly exciting given the absence of any such speculation in the existing literature. However, the fact that such features are known to exist in hydrology and other related fields makes this discovery somewhat less surprising. Connections of these features with the underlying mechanism of seismic noise generation are certainly in the calling and reserved for further study. Based on the statistical characterization, we developed a composite linear time-series (ARIMA-GARCH) model for the background noise. The ARIMA component of the model is capable of modeling the integrating type (but not the conditional heteroskedasticity) and linear correlation in noise, while the GARCH component of the model captures the conditional heteroskedasticity of the data. It may also be emphasized that the ARIMA part is built on the data while the GARCH component is developed from the residuals of the ARIMA model.

Variation in these properties, and hence the uniqueness and uniformity of the model structure, are studied by analyzing seismic noise for a fixed location for daily and monthly variation. Model structure is consistent with time and climatic conditions, however, estimated parameters vary with time. The scope of this work has been limited to linear models. However, this restriction does not dilute the significance of our findings in any discernible way. Future work consists of investigation into the physical reasoning of the non-stationary and non-linear nature of seismic noise, whenever applicable. Further, the improvement that the new class of models can potentially have in applications such as detection of P-wave onset is a natural next step in the course of seismic analysis. In closing, we believe that the findings of this work can lead to a more sophisticated and efficient algorithm for onset detection.


\appendix
\section*{Appendix}
\label{sec:Appendix}
\section{Testing for integrating effects}\label{sec:ADFandPP}
Statistical tests such as ADF (\cite{said1984testing}) and PP (\cite{phillips1988testing}) test are used for testing the presence of integrating effects in correlated data. Integrating effects are said to be present whenever processes (signals) have a long memory or very low-frequency characteristics, one or both of which seismic waves can be intuitively expected to possess. Random walk processes are considered to be non-stationary because they do not satisfy the conditions of weak (second-order) stationarity (\cite{shumway2017time}). In fact, integrating effects are modeled as auto-regressive processes of first-order with the pole at unity. An AR process with only one pole at unity (with other poles inside the unit circle) is termed as an $I(1)$ process. For a given time-series, $y[k] = \{y[1],y[2], \ldots, y[N]\}$, both ADF and PP tests assess the null hypothesis that $y[k]$ is an $I(1)$ process against the alternate of $I(0)$ (no pole at unit circle) using the model given by:
\begin{align}
\label{eq:ADF}
y[k] =& \rho y[k-1] + u[k]
\end{align}
where, $u[k]$ is a stationary process modeled as 
\begin{align}
u[k] =& \sum_{i=1}^{p}\alpha_{i}u[k-i] + \sum_{j=1}^{m}\beta_{j}e[k-j] + e[k] \label{eq:adf_2}
\end{align}
The parameters $p$ and $m$ are the AR and MA orders respectively, while $e[k]$ is the driving white-noise process (further details are provided in the ensuing paragraph). The unit root hypothesis test is now formulated as:
\begin{align*}
H_0: \rho = 1 \\
H_a: \rho < 1
\end{align*}
where all the parameters in \eqref{eq:ADF} and \eqref{eq:adf_2} are estimated using least square regression in both tests. 

Both the tests differ in the way they exploit serial correlation in $u[k]$ and the assumptions they make on the innovation process $e[k]$. On one hand, the ADF test assumes $e[k]$ to be \textit{i.i.d.} \textit{homoskedastic} process and accounts for the serial correlation in $u[k]$ by approximating it as higher order AR process in the form of lagged differenced regressors of data ($\Delta y[k-1], \Delta y[k-2],...$, where $\Delta$ is the differencing operator such that $\Delta y[k] = y[k]-y[k-1]$). On the other hand, the PP test can handle a broader class of situations involving \textit{unconditionally heteroskedastic} innovation process. Moreover, it uses a non-parametric adjustment in test statistic to account for the serial correlation in $u[k]$, making it more robust.

\section{Testing for heteroskedasticity}\label{sec:PSRtest}
Depending on the nature of heteroskedasticity, unconditional or conditional, different tests are implemented to identify the respective heteroskedastic behaviour of a time-series. One of the widely used tests for heteroskedasticity of a time-series is the PSR test (\cite{priestley1969test}) which is based on the uniformity of time-varying spectral density, known as \emph{evolutionary spectral density}, evaluated at different instants in time. The test is devised for detecting overall stationarity of the second-order properties, i.e., it tests for non-stationarities in variance as well as in spectral properties. A primary assumption underlying the PSR test is that the time-series is first-order stationary.

For a time-series $y[k]$ with evolutionary spectral density function $f_{k}(\omega)$, where $\omega$ is the frequency, the test checks how non-constant $f_{k}(\omega)$ is as a function of time $k$. The test works with logarithmic transformation of estimated evolutionary spectral density which is defined as,
\begin{align}
Y(k,\omega) = \log(\hat{f}_{k}(\omega))
\end{align}
where, $\hat{f}_{k}(\omega)$ is the estimate of evolutionary spectral density of the series $y[k]$. Then, approximately (\cite{priestley1969test}),
\begin{align*}
E(Y(k,\omega)) &= \log(f_{k}(\omega)) \\
\text{var}(Y(k,\omega)) &= \sigma^{2}
\end{align*}
where, $\sigma^{2}$ is independent of $k$ and $\omega$. Logarithmic transformation diverts the focus from variation in second-order properties to the changes in mean structure of $Y(k,\omega)$.

The test assesses the null hypothesis of homoskedasticity (constant variance) against the alternate of heteroskedasticity. It is a right-tailed test, i.e., the null hypothesis is rejected in favour of alternate hypothesis if the test statistic is greater than the critical value for a chosen significance level.

Another most widely used tests for \emph{conditional} heteroskedasticity of a time-series is \textit{ARCH test} (\cite{engle1982autoregressive}) which is implemented on the residual series obtained from a time-series model. The basic idea is to fit a linear regression model to the squared residuals and examine whether the fitted model is significant or not. The test assesses the null hypothesis that series of residuals $X[k]$ exhibit no conditional heteroskedasticity against the alternative that an ARCH model can explain the underlying process using the model given by:
\begin{align}
X^{2}[k] =& \gamma_{0} + \gamma_{1}X^{2}[k-1] + \gamma_{L}X^{2}[k-L] + \epsilon_{r}[k] 
\end{align}
where, $L$ is the ARCH order, $\gamma_{i}$ are the model parameters, and  $\epsilon_{r}[k] \sim$ GWN $(0,1)$. Hypothesis for presence of the ARCH effect is formulated as:
\begin{align*}
H_{0}: & ~ \gamma_{0} = \gamma_{1} = \ldots = \gamma_{L} = 0\\
H_{1}: & ~ \exists \gamma_{i} \neq 0
\end{align*}
The test statistic for ARCH test is the Lagrange multiplier and has an asymptotic chi-square distribution with $L$ degrees of freedom under the null hypothesis.

\section{Testing for Gaussianity}\label{sec:GaussianityTest}
A time-series $y[k]$ is said to be Gaussian random process if, for all $ k_{1},k_{2},\ldots,k_{n}$, the random variables $y[k_{1}],y[k_{2}],\ldots,y[k_{n}]$ are jointly normal. Shapiro-Wilk test (\cite{shapiro1965analysis}) is one of the most powerful and widely used test for testing normality of a time-series. The test assesses the null hypothesis that data is generated from a Gaussian process with unknown parameters ($\mu$, $\sigma^2$). Given a time-series, $y[k]=\{y[1],y[2],\ldots,y[N]\}$, the test statistic for Shapiro-Wilk test is defined as
\begin{align}
W = \dfrac{\bigg(\sum\limits_{j = 1}^{N}a_{j}y_{(j)}\bigg)^{2}}{\bigg(\sum\limits_{j = 1}^{N}(y[j]-\bar{y})^{2}\bigg)}
\label{eq:SW}
\end{align}
where, $y_{(j)}$ represents the \emph{order statistic} of $y[k]$ which is defined as $y_{(1)}\leq y_{(2)}\leq \ldots \leq y_{(N)}$. The weighting coefficients $a_{j}$ are derived from the first and second-order properties of the order statistic of a standard Gaussian white-noise process ($z[k] \sim \text{GWN}(0,1)$) of size $N$ and are defined as:
\begin{align}
\mathbf{a'} = (a_{1},a_{2},\ldots,a_{N}) = \dfrac{\mathbf{m'}\mathbf{V}^{-1}}{(\mathbf{m'}\mathbf{V}^{-1}\mathbf{V}^{-1}\mathbf{m})}
\end{align}
where, $\mathbf{m'} = (m_{1},m_{2},\ldots,m_{N})$ are the expected values and $\mathbf{V}$ is the variance-covariance matrix of order statistic of $z[k]$. For different sample size, \cite{shapiro1965analysis,royston1982extension,royston1992approximating} provide tables to compute the weighting coefficients and p-value for a chosen significance level. Values of $W$ are bounded between $0$ and $1$. For values of $W$ close to $1$, null hypothesis of normality cannot be rejected. On the other hand, smaller values of $W$ indicate departure from normality.

\section{Testing for linearity}
\label{sec:LinearityTest} A stationary random process is said to be linear if and only if it can be represented as
\begin{align}
y[k] = \sum_{n=-\infty}^{\infty}h_{n}e[k-n], \quad \forall k
\label{eq:linear}
\end{align}
where, $e[k]\sim \text{GWN}(0,1)$ and the weights $h_{n}$ satisfy $\sum\limits_{n=-\infty}^{\infty}|h_{n}| < \infty$. The condition on  weights guaranteeing stationarity of $y[k]$ also implies absolute convergence of the auto-covariance function (of $y[k]$). Among the several methods available for testing linearity,the surrogate-based approach has emerged as a powerful method, especially in the recent years. It offers two major advantages over the traditional nonlinearity tests such as time-reversibility test, BDS test, etc., (\cite{bisaglia2014testing,tsay1986nonlinearity}). The two benefits are (i) the test can be tailored for a specific null hypothesis and (ii) it is compatible with any test statistic that can be selected independently of the null hypothesis. The performance of surrogate-based test for linearity depends on the algorithm used for generating surrogates and the discriminating test statistic that is deployed. The null and alternative hypotheses for linearity test are:
\begin{align*}
H_{0}:& \text{ Data is generated from a linear}\\
&\text{Gaussian process}\\
H_{a}: & \text{ Data is nonlinear}
\end{align*}
where the correlation dimension (to be defined shortly in \eqref{eq:corrdim}) is used as the test statistic for testing linearity. At a chosen significance level, if the correlation dimension of data does not lie within the lower and upper critical values of the test statistic for surrogates, then $H_{0}$ is rejected in favour of alternate hypothesis. We next describe the method for generating surrogates and provide the definition of correlation dimension.

\subsection{Surrogate data generation \label{subsec:surrogates}}
The concept of surrogate data was first introduced in the field of physics by \cite{theiler1992testing} to detect non-linear structures in the stationary time-series. Surrogates or virtual realizations are generated either using an exact model structure with fixed parameters (known as \emph{typical realizations}) or through random realizations which preserves certain properties of data (known as \emph{constrained realization}). The former approach is restricted to specific type of processes as surrogates are generated using the estimated model from data while the latter approach is more flexible and can be used for a wide range of processes (\cite{theiler1996constrained}).

Random realizations are generated through a procedure known as randomization, while preserving certain properties of data that are commensurate with the null hypothesis. There exists different ways of randomizing the data, namely, phase randomization (also known as Fourier transform (FT) surrogates), amplitude adjusted Fourier transform (AAFT), iterative AAFT, etc., (D.\cite{kugiumtzis2002surrogate}, \cite{theiler1992testing}). FT surrogates are generated by randomizing the phase of Fourier coefficients of data followed by inverse Fourier transform (assuming that amplitude distribution of data is normal). These surrogates preserve the original linear correlation for sufficiently large lags. For most real-time applications, data fails to follow the assumption of normal distribution. In such cases, FT surrogate method results in erroneous implementation of the linearity test. In contrast, surrogates generated from the AAFT method not only preserves the linear correlation but also the marginal cumulative density function (CDF) of data. There are three steps to generating surrogates using AAFT, (i) \textit{Gaussianization} transform of data to obtain normal distribution, (ii) \textit{surrogate generation} using FT method and (iii) \textit{inverse Gaussianization} transform to regain the original marginal CDF. 
The algorithm is known to be sensitive to the sample size resulting in bias for small sample data (\cite{SCHREIBER2000346}).

\subsection{Correlation dimension ($D_{2}$)}\label{subsec:CorrelationDim}
Correlation dimension, introduced by \cite{grassberger1983measuring} in the field of chaos theory, is a member of fractal dimension family which is widely used to test nonlinearity in data. It is a measure of dimensionality, of the space occupied by the random data points, which indicates the minimum number of variable required to model the behaviour of system in phase space. For an $m$-dimensional phase space, the correlation integral $C(r)$ is defined as:
\begin{equation}
    C(r) = \lim\limits_{N\to \infty}\dfrac{2}{N(N-1)}\sum\limits_{i,j} H(r - |y_{i} - y_{j}|)
\end{equation}
where, $H$ is the Heaviside step function, $r$ is the Euclidean distance between the embedded points in the phase space, $N$ is the length of the data $\mathbf{y}$. The parameter $D_{2}$ is the slope of log-log graph of $C(r)$ versus $r$,
\begin{align}
D_{2} = \lim\limits_{r\to 0}\frac{\log(C(r))}{\log(r)}
\label{eq:corrdim}
\end{align}
The quantity $C(r)$ quantifies the number of pair of points which have a distance less than or equal to $r$.

\section{Time-series models}\label{sec:tsmodel}
A linear stationary random process is said to be generated by an ARIMA($p,d,m$) model, if it can be represented by the following difference equation:

{\footnotesize \begin{align}
	\Big(1-\sum_{i=1}^{p}\phi_{i}q^{-i}\Big)(1-q^{-1})^{d}y[k] = \Big(1 + \sum_{j = 1}^{m}\theta_{j}q^{-j}\Big)e[k]
	\label{eq:tsmodel}
	\end{align}}
\noindent where, $q^{-1}$ is the backshift operator, $\phi_{i}$ and $\theta_{j}$ are the AR and MA coefficients of order $p$ and $m$ respectively, $d$ is the degree of differencing and innovations $e[k]\sim \text{GWN}(0,\sigma_{e}^{2})$. The ARIMA model essentially captures the integrating effects by constructing an ARMA representation on differenced data. Estimation of an ARIMA model  (\cite{chatfield2016analysis,box2015time}) involves identifying the values of $d,p,m$, estimating the unknown parameters of \eqref{eq:tsmodel} and $\sigma_{e}^{2}$ using the estimation algorithms such as least squares, maximum likelihood, etc.

The ARIMA class of time-series models fail to model heteroskedastic processes. For processes that are \emph{unconditionally} heteroskedastic, the standard remedy is to apply a suitable transformation such as Box-Cox (\cite{sakia1992box}) to the data prior to building an ARIMA model. On the other hand, for processes that are \emph{conditionally} heteroskedastic, ARIMA models can be developed but result in prediction errors that are linearly independent (uncorrelated) whereas the squared residuals are correlated, i.e., a non-linear dependence of specific nature exists among the residuals. This unusual characteristic of prediction errors can be nicely explained using ARCH / GARCH models.

\subsection{Generalized ARCH (GARCH) models}\label{subsec:ARCHmodel}
ARCH models or their generalized versions (GARCH) provide the framework to explicitly model the time-varying variance in mean stationary uncorrelated time-series. These class of models were introduced in the econometrics and finance applications by \cite{engle1982autoregressive} and \cite{bollerslev1986generalized} respectively to model conditional heteroskedasticity in wage-price data. A generalized-ARCH (GARCH) model of order ($P,Q$) is defined as:
\begin{align}
&X[k] = \sigma_{k}\epsilon[k]\\
\sigma_{k}^{2} &= c_{0} + \sum_{i = 1}^{P}b_{i}X_{k-i}^2 + \sum_{j = 1}^{Q}a_{j}\sigma_{k-j}^{2}
\label{eq:ARCH}
\end{align}
where $P (\geq 1)$ and $Q(\geq 0)$, while $c_{0} \geq 0,~ b_{i}\geq 0, ~ a_{j} \geq 0$ are constants. The driving force $\epsilon[k] $ is i.i.d.$(0,1)$ and independent of $X_{k-l}, l\geq 1$ for all $k$. One can observe from \eqref{eq:ARCH} that the GARCH model is essentially an ARMA representation for $\sigma^2_k$ in terms of prediction errors and the variance of prediction error.

In general, GARCH models are developed on residuals obtained from the ARIMA model for the original series. Thus, the series $X[k]$ in \eqref{eq:ARCH} is the residual obtained from an optimally estimated ARIMA($p,d,m$) model for the given series $y[k]$. The resulting composite model for $y[k]$ is known as an ARIMA($p,d,m$)-GARCH($P,Q$) model.

\end{document}